\documentclass[useAMS,usenatbib]{mn2e}

\usepackage{graphicx}
\usepackage{amsmath}
\usepackage{amssymb}
\usepackage{appendix}
\usepackage{color}
\usepackage{soul}

\def\be{\begin{equation}}
\def\ee{\end{equation}}
\def\ba{\begin{eqnarray}}
\def\ea{\end{eqnarray}}
\def\go{\mathrel{\raise.3ex\hbox{$>$}\mkern-14mu
             \lower0.6ex\hbox{$\sim$}}}
\def\lo{\mathrel{\raise.3ex\hbox{$<$}\mkern-14mu
             \lower0.6ex\hbox{$\sim$}}}

\newcommand{\hatl}{\hat{\mbox{\boldmath $l$}}}
\newcommand{\tildel}{\tilde{\mbox{\boldmath $l$}}}
\newcommand{\Gv}{\mbox{\boldmath $G$}}

\begin{document}

\title[Evolution of linear warps in accretion discs]
{Evolution of linear warps in accretion discs 
and applications to protoplanetary discs in binaries}
\author[F. Foucart and D. Lai]
{Francois Foucart$^{1}$\thanks{Email: ffoucart@cita.utoronto.edu}
and Dong Lai$^{2}$\\
$^1$Canadian Institute for Theoretical Astrophysics,
  University of Toronto, Toronto, Ontario M5S 3H8, Canada\\
$^2$Department of Astronomy, Cornell University, Ithaca, NY
  14853, USA}

\pagerange{\pageref{firstpage}--\pageref{lastpage}} \pubyear{2010}

\label{firstpage}

\maketitle

\begin{abstract}
Warped accretion discs are expected in many protostellar binary systems.
In this paper, we study the long-term evolution of disc warp and precession for discs with dimensionless thickness $H/r$ larger than their viscosity parameter $\alpha$,
such that bending waves can propagate and dominate the warp evolution.
For small warps, these discs undergo approximately rigid-body precession. We derive analytical expressions for the warp/twist profiles of the disc and the alignment timescale for a variety of models. Applying our results to circumbinary discs, we find that these discs align with the orbital plane of the binary on a timescale comparable to the global precession time of the disc, and typically much smaller than its viscous timescale. We discuss the implications of our finding for the observations of misaligned circumbinary discs (such as KH 15D) and circumbinary planetary systems (such as Kepler-413); these observed misalignments provide useful constraints on the uncertain aspects of the disc warp theory. On the other hand, we find that circumstellar discs can maintain large misalignments with respect to the plane of the binary companion over their entire lifetime. We estimate that inclination angles larger than $\sim 20^\circ$ can be maintained for typical disc parameters. Overall, our results suggest that while highly misaligned circumstellar discs in binaries are expected to be common, such misalignments should be rare for circumbinary discs. These expectations are consistent with current observations of protoplanetary discs and exoplanets in binaries, and can be tested with future observations. 
\end{abstract}

\begin{keywords}
accretion, accretion discs -- hydrodynamics -- planetary systems: 
protoplanetary discs -- stars: binary
\end{keywords}

\section{Introduction}

Warped accretion discs are expected in a variety of astrophysical
systems, such as protoplanetary discs, discs around spinning black
holes, remnant discs following neutron star binary mergers and stellar
tidal disruption by massive black holes. In particular, in
proto-stellar binary systems, both circumstellar discs and
circumbinary discs are likely formed with inclined orientations, as a
consequence of the complex star/binary/disc formation processes 
(e.g., Bate et al.~2003; McKee \& Ostriker 2007; Klessen 2011).
Indeed, a number of binary young stellar objects (YSOs)
are observed to contain circumstellar discs that are misaligned with
the binary orbital plane (e.g., Stapelfeldt et al.~1998).
Observations of jets along different directions in unresolved YSOs
also suggest the existence of misaligned discs (e.g., Davis, Mundt \&
Eisl\"offel 1994; Roccatagliata et al.~2011).  Additionally, imaging
of circumbinary debris discs show that while the disc plane and the
binary orbital plane are aligned for some systems (such as $\alpha$
CrB, $\beta$ Tri and HD 98800), significant misalignments can occur in
others (such as 99 Herculis, with mutual inclination $\go 30^\circ$;
see Kennedy et al.~2012a,b). Finally, the pre-main sequence binary
KH~15D is surrounded by a precessing circumbinary disc inclined with
respect to the binary plane by $10^\circ$-$20^\circ$ (e.g., Winn et
al.~2004; Chiang \& Murray-Clay 2004; Capelo et al.~2012), while in
the FS Tauri system, circumbinary and circumstellar discs appear to be
misaligned (Hioki et al.~2011).

Also of relevance is the recent discovery of transiting planetary
systems around stellar binaries (e.g., Doyle et al.~2011; Welsh et
al.~2012; Orosz et al.~2012; Kostov et al.~2014).  While most of these
systems have the binary orbital plane and the planetary orbital plane
consistent with alignment, a small misalignment ($\sim 2.5^\circ$) has
been found in the Kepler-413 system (consisting of a Neptune-size
planet orbiting an eclipsing stellar binary; Kostov et al.~2014).  A
potential explanation for this misalignment is that the planet formed
in a circumbinary disc misaligned with respect to its host binary.
Understanding the inclination distribution of circumbinary planets 
is important for determining the abundance and the formation scenarios
of these planets (e.g., Martin \& Triaud 2014; Armstrong et al.~2014).

Theoretical studies of the evolution of misaligned accretion discs
have been performed under the assumption of small warps in the discs,
with numerical simulations providing information for the behavior of
discs with larger warps.  For 
discs in which the
viscosity parameter $\alpha$ is smaller than $H/r$ (with $H$ the
scaleheight of the disc),
bending waves can propagate warps through
the disc at about half the disc sound speed (
Papaloizou \& Lin 1995). This rapid communication allows a
misaligned disc experiencing an external torque to precess
approximately as a rigid body, as long as the precession timescale of
the disc is longer that the time required for bending waves to travel
across the disc (Papaloizou \& Terquem 1995).  This analytical
prediction has been verified by 3D numerical simulations of
circumstellar discs using both smoothed particle hydrodynamics
(Larwood \& Papaloizou 1997) and finite volume methods (Fragner \&
Nelson 2010).  Smoothed particle simulations of narrow circumbinary
discs have provided similar results (Facchini et al.~2013). Recent
numerical simulations have also
been used to explore the limitations
of the linear theory (Fragner \& Nelson 2010; Sorathia et al. 2013a).

A one-dimensional model for the evolution of small warps in a thick
disc has been developed by Lubow \& Ogilvie 2000 (see also Lubow
et al.~2002, Ogilvie 2006), and used to study the evolution of
circumstellar discs. In particular, Lubow \& Ogilvie (2000) used this
model to study misaligned circumstellar discs in binaries, 
and compute discrete eigenmodes of the warp.
They found that, for specific values of the outer radius
of the disc, unstable modes exist which can drive the growth of the
disc inclination. Away from these resonant radii, the
inclination of the disc is damped on a timescale which can be computed
numerically by solving an eigenvalue problem. An analytical 
estimate
for the damping rate of the inclination of circumstellar disc has also
been proposed by Bate et al.~(2000), based on 
a global estimate of the energy dissipation
in the disc due to viscous torques. This damping
timescale is generally of the order of the viscous timescale of the
disc. However, Gammie et al.~(2000) found that 
even when the disc warp is small enough to satisfy the linear theory,
the disc may still be susceptible
to parametric instabilities, whose effect on the long term evolution
of the disc is currently uncertain. Bate et al.~(2000) postulate that
parametric instabilities will cause the damping of the inclination on
a timescale comparable to the growth timescale of the instability,
and would cause circumstellar discs to rapidly align within an angle
of the order of the disc thickness $H/r$. Estimates of the condition
under which parametric instabilities can grow, and of their growth
rate have more recently been improved by the work of Ogilvie \& Latter
(2013),

Some of the results obtained for circumstellar discs 
can be applied
to the less studied circumbinary discs. The condition and
the timescale for global precession (Papaloizou \& Terquem 1995) 
can be similarly obtained, 
and the approximate formula for the damping of the
disc inclination proposed by Bate et al.~(2000) 
appears to match the results of one-dimensional 
simulations performed by Facchini et al.~(2013) to within
factors of a few. The disc profile and the alignment of the
disc-binary system were also studied in the limit of infinite
circumbinary discs by Foucart \& Lai (2013), 
and the evolution of finite size discs towards an equilibrium
warped profile has been demonstrated numerically
by Facchini et al.~(2014).

In this paper, we study in more details the warp profile of both
circumstellar and finite circumbinary discs. Building on the methods
of Papaloizou \& Terquem (1995) and on the eigenmode decomposition
performed by Lubow \& Ogilvie (2000), we provide approximate analytical
expressions for the orientation profile of the disc and the damping
timescale of the inclination by considering the properties of the
lowest order-mode. 
With the warping profile closest to a flat disc, such a low-order mode 
has the smallest internal stresses and the longest lifetime, and will
dominate the long term evolution of the disc. We show that these
approximate formulae are correct to first order in the disc warp,
and provide solutions which match well the results obtained both
by solving the eigenvalue problem exactly, and by full 3D
simulations. In particular, our results allow for the computation of the
damping timescale of the 
disc inclination (both for circumstellar discs with an external binary
companion and for circumbinary discs),
which is much more accurate than the formula previously 
derived by Bate et al.~(2000), without the need for
expensive simulations or the explicit 
solution of an eigenvalue problem.
Because we directly recover the orientation of the disc at
all radii, we can also easily compare our disc warp profiles with the
predictions of Ogilvie \& Latter (2013) for the local conditions leading
to the growth of parametric instabilities, and thus obtain improved
estimates on the conditions under which these instabilities might
affect the evolution of the disc.  We show that
circumstellar discs are likely to be able to maintain significant
misalignments with respect to the orbital plane of a binary companion,
in contrast to the early conclusion of Bate et al.~(2000).

In Sec.~\ref{sec:model} we present our analytical model for disc warps
and inclination damping, and compare our results with numerical solutions.
We then apply our calculations to the study of circumbinary discs in 
Sec.~\ref{sec:circum}, and circumstellar discs in Sec.~\ref{sec:circumstellar}.
We consider the effects of parametric instabilities in Sec.~\ref{sec:param}.
In Sec.~\ref{sec:kh} we revisit the case of the KH 15D
system, and examine how our results impact the interpretation of that
system as a truncated circumbinary disc (Winn et al.~2004,
Chiang \& Murray-Clay~2004, Lodatto \& Facchini 2013).
In Sec.~7 we summarize our results and discuss their implications.

\section{disc warp and Inclination Damping: Theoretical model and Analytical Solutions}
\label{sec:model}

We consider the evolution of an accretion disc in the gravitational potential
\be
\Phi({\bf r}) = -\frac{GM}{r} + \delta \Phi(r,z),
\ee
assuming that the deviation $\delta \Phi$ from the pure Keplerian potential 
$(-GM/r)$ is small (in a sense to be specified below). 
For circumbinary discs, we have $M=M_1+M_2$ with $M_{1,2}$ the masses
of the two binary components; for circumstellar discs, we have
$M=M_\star$ with $M_\star$ the mass of the central star.  The
potential $\delta\Phi(r,z)$ is axisymmetric, with the coordinate $z$ along 
its symmetry axis, the angular momentum axis of the binary. 
We also assume that $\delta\Phi$ is either time-independent or time-averaged.
The disc is generally misaligned with respect to the $z$-axis, with
the angular momentum of the disc element at radius $r$ lying at time
$t$ in the direction $\hatl(r,t)$. In cartesian coordinates, we write
\be
\hatl(r,t)=(\sin{\beta}\cos{\phi},\sin{\beta}\sin{\phi},\cos{\beta}),
\ee
where $\beta (r,t)$ is the warp angle (the inclination of $\hatl$
relative to the $z$-axis), and $\phi(r,t)$ is the twist angle (corresponding to
the rotation of $\hatl$ around the $z$-axis).

We consider a vertically integrated model for the disc, with surface density
$\Sigma$ and integrated pressure given by
\be
\Sigma = \int\!\rho\, dz,\quad
P = \int\!p\,dz=\Omega_z^2 \Sigma H^2. \label{eq:H}
\ee
Here $\rho$ is the density in the disc, $p$ the pressure, $\Omega_z$
the frequency for vertical oscillations in the potential $\Phi$, and
$H$ is an effective scale height defined by Eq.~(\ref{eq:H}). We also
define $c_s = H \Omega_z$. For a vertically isothermal disc, $c_s$ is
the isothermal sound speed and $H$ the usual Gaussian scale height
(see Lubow et al. 2002).
If the disc warp is linear, i.e., if the variation of the disc orientation 
satisfies $\psi\equiv |\partial \hatl/\partial \ln{r}| \ll 1$, theoretical studies
of $\alpha$-discs (with isotropic viscosity $\nu=\alpha H^2 \Omega$,
where $\alpha$ is the Shakura-Sunyaev parameter, and
$\Omega$ is the angular velocity) have shown that the evolution of the
warp follow two possible behaviors (Papaloizou \& Pringle 1983;
Papaloizou \& Lin 1995). 
For discs with $H/r\lo\alpha$, the warp
satisfies a diffusion-type equation with the diffusion coefficient $\nu_2
= \nu/(2\alpha^2)$,
at least for $\alpha \ll 1$ (see Ogilvie 1999 and Lodato \& Price 2010 for
analytical and numerical studies of the effects of a larger $\alpha$). 
For thicker discs ($H/r\go\alpha$), the warp propagates
as bending waves at about half the sound speed $c_s$ and dams
on a timescale $1/(\alpha\Omega)$.
In the latter case, the numerical simulations of Sorathia et al. (2013a)
suggest that the linear theory remains valid for small warp satisfying
$\psi\equiv |\partial \hatl/\partial \ln{r}| \leq H/r$ while for stronger warps nonlinear
effects cause rapid damping of the inclination of the disc. For larger viscosities,
bending waves damp on a length scale shorter than the scale height of the disc,
and do not propagate across the disc.
In this paper,
we will consider discs with $H/r\go\alpha$,
which are relevant for a large class of 
astrophysical systems including protoplanetary discs.
%

For discs with $H/r\go \alpha$
and linear warps, the orientation of the disc 
evolves according to (Lubow \& Ogilvie 2000; see also Lubow et al.~2002 and Ogilvie 2006) 
\ba 
&&\Sigma r^2 \Omega
\frac{\partial \hatl}{\partial t} = \frac{1}{r} \frac{\partial
  \Gv}{\partial r} + \Sigma r^2 \Omega^2 \left(\frac{\Omega^2 -
  \Omega_z^2}{2\Omega^2}\right) \hat{\bf e}_z \times \hatl , \label{dldt}\\ 
&&\frac{\partial \Gv}{\partial t} = \frac{\Sigma
  c_s^2 r^3 \Omega}{4} \frac{\partial \hatl}{\partial r} +
\left(\frac{\Omega^2-\kappa^2}{2\Omega^2}\right) 
\Omega
\hat{\bf e}_z \times
\Gv - \alpha \Omega \Gv , \label{dGdt} 
\ea 
where $\hat{\bf e}_z$ is the unit vector along the $z$-axis,
${\bf G}$ is the internal stress in the disc, and $\Omega_z$ and $\kappa$ are the
vertical and radial epicyclic oscillation frequencies.
The first term on the right-hand side of
Eqs.~(\ref{dldt})-(\ref{dGdt}) describe the free propagation of
bending waves in the disc. The last term in Eq.~(\ref{dldt})
corresponds to the external torque that the potential $\delta\Phi$
applies on the misaligned disc. The last two terms in Eq.~(\ref{dGdt})
arise from periastron advance for non-Keplerian orbits and from
viscous damping in the disc, respectively.
For small misalignments between $\hatl$ and 
$\hat{\bf e}_z$, 
the frequencies
$\Omega_z$ and $\kappa$ can be computed according to 
\ba 
\kappa^2 &=& \frac{2\Omega}{r} \frac{d (r^2\Omega)}{dr}, \label{eq:kappa}\\ 
\Omega_z^2 &=&\left(\frac{\partial^2\Phi}{\partial z^2}\right)_{z=0}\label{eq:omegaz}.
\ea
The validity of Eqs.~(\ref{dldt})-(\ref{dGdt}) requires a small
deviation from the Keplerian potential, i.e., $|\kappa^2 -\Omega^2|
\lo \delta \Omega^2$ and $|\Omega_z^2-\Omega^2| \lo \delta \Omega^2$,
where $\delta=H/r$.  Under these assumptions,
Eqs.~(\ref{dldt}-\ref{dGdt}) have been shown to reproduce the results
of 3D numerical simulations for the evolution of linear warps (see,
e.g., Larwood \& Papaloizou 1997 for circumstellar discs, or Facchini
et al.~2013 for circumbinary discs). 
For large misalignments between $\hatl$ and $\hat{\bf e}_z$, Eqs.~(\ref{eq:kappa}-\ref{eq:omegaz})
would have to be corrected to appropriately reflect the time-averaged effects of the gravitational
potential on the warp in a plane different from $z=0$, but the general properties of the equations
are left unchanged. This should be contrasted with the requirements of a quasi-Keplerian potential
and of a small warp $|\partial \hatl/\partial \ln{r}|\ll 1$, which are both necessary for the linear theory
of bending wave propagation to apply.

There are however a couple of important caveats to keep in mind when
using this one-dimensional model for 
the warp evolution.
The first is that it was derived under the assumption
of isotropic viscosity $\nu=\alpha H^2 \Omega$. 
In general, the viscosity may be anisotropic, and there is
thus no guarantee that, in such case, the viscosity
parameter $\alpha$ entering Eq.~(\ref{dGdt}) is identical to the
viscosity parameter responsible for the radial transfer of mass and
angular momentum. 
Shearing box simulations of turbulence have found reasonable agreement with
the assumption of isotropic viscosity (Torkelsson et al. 2000), 
at least as far as its influence on the damping parameter $\alpha$ is concerned. 
But recent
global simulations find that the viscosity generated by the magnetorotational
instability is anisotropic (Sorathia et al. 2013b, Teixeira et al. 2014).
In practice, we can take $\alpha$ as a simplified parametrization of all 
the physical effects causing a damping of bending waves in the disc.

The second caveat is that, even in the linear regime, the disc warp may be
unstable to a parametric instability and the development of turbulence
due to the strong vertical shear in the flow velocity (Gammie et al.~2000; 
Ogilvie \& Latter 2013).
When this occurs, the behavior of the disc warp is uncertain.  One
reasonable proposal is that the damping of the disc warp would then be
set by the growth rate of the instability (Bate et al.~2000). For a
given disc profile, the parametric instability would then effectively
set a minimum value for $\alpha$. We discuss this issue in more
detail in Sec.~\ref{sec:param}.

Despite these limitations, the linear model for the propagation of
bending waves remains a useful
tool to obtain approximate solutions to the evolution of warped discs,
without having to resort to expensive numerical simulations that
include all the microphysical effects required to recover the exact
form of the viscosity in the disc. In the following, we derive 
analytical approximations for the main properties of the disc (warp amplitude,
internal stress, global precession frequency, damping timescale of the
warp) for configurations in which the disc, as a result of external
torques, 
undergoes an approximately solid-body precession.
We then compare these analytical
results with numerical solutions in Sec.~\ref{sec:numsol}.

\subsection{Warp equations as an eigenvalue problem}

For time-independent (or time-averaged) potentials $\Phi$, the
solution of the evolution equations for $\hatl$ and $\Gv$ can be
written as a sum of eigenmodes
\ba 
&&\hatl(r,t) = \Re{\left(\sum_\omega \tildel_\omega(r)
  e^{-i\omega t}\right)} ,\\ 
&&\Gv(r,t) =\Re{\left(\sum_\omega \tilde{\Gv}_\omega(r) e^{-i\omega t}
  \right)} ,
\ea 
where $\omega$ is the complex eigenfrequency of the mode, and
$(\tildel_\omega,\tilde{\Gv}_\omega)$ are complex functions of
the radius $r$. Physically, the real part of $\omega$ corresponds to the
global precession of the angular momentum of the disc around the
symmetry axis of the gravitational potential if 
$\tildel_{\omega,y}= \pm i\tildel_{\omega,x}$ 
(which, as we will see below, is always the case).
Its imaginary part corresponds to the exponential damping of the warp (or
exponential growth, if the mode is unstable). Global precession is
possible as long as the small warp condition $|\partial \hatl/\partial
\ln{r}| \ll 1$ is satisfied. Papaloizou \& Terquem (1995) showed that
this is roughly equivalent to the condition
\be
\Re{(\omega)} \lesssim \frac{H}{r} \Omega(r_{\rm out}),
\ee
which can be rephrased as the condition that the precession timescale
is
longer that the travel time of bending waves across the disc
(as bending waves travel at half the sound speed $c_s\sim H\Omega$).

Each eigenmode then satisfies a set of four complex ODEs: 
\ba
 \frac{d \tilde{G}_x}{dr}&=&\Sigma r^3 \Omega \left(-i \omega \tilde{l}_x + Z(r)\Omega \tilde{l}_y\right)\label{eq:Gx}\\
 \frac{d \tilde{G}_y}{dr}&=&\Sigma r^3 \Omega \left(-i \omega \tilde{l}_y - Z(r)\Omega \tilde{l}_x \right) \label{eq:Gy}\\
 \frac{\Sigma c_s^2 r^3 \Omega}{4} \frac{d \tilde{l}_x}{dr}&=&(-i\omega +\alpha \Omega)\tilde{G}_x + K(r)\Omega \tilde{G}_y \label{eq:lx}\\
 \frac{\Sigma c_s^2 r^3 \Omega}{4} \frac{d \tilde{l}_y}{dr}&=&(-i\omega +\alpha \Omega)\tilde{G}_y - K(r)\Omega \tilde{G}_x\label{eq:ly},
\ea
where we have defined 
\ba
&& K(r) = \frac{\Omega^2-\kappa^2}{2\Omega^2},\label{eq:K(r)}\\
&& Z(r) = \frac{\Omega^2-\Omega_z^2}{2\Omega^2},\label{eq:Z(r)}
\ea
and dropped the subscripts $\omega$ for convenience.
These equations can be solved as an eigenvalue problem once proper
boundary conditions are chosen at the inner and outer edges of
the disc. For the finite size discs considered in this paper, we will adopt
the zero-torque boundary condition: $G(r_{\rm in})=G(r_{\rm out})=0$.

The four ODEs (\ref{eq:Gx})-(\ref{eq:ly}) can easily be reduced to two
sets of ODEs which are only coupled by sharing the same eigenfrequency
$\omega$. Defining $W_\pm = \tilde l_x \pm i \tilde l_y$ and
$G_\pm = \tilde G_x \pm i \tilde G_y$, we find (see Lubow et al.~2002)
\ba
\frac{dG_\pm}{dr} &=& -i \Sigma r^3 \Omega W_\pm (\omega \pm Z \Omega)\label{eq:GLO}\\
\frac{dW_\pm}{dr}&=& \frac{4G_\pm}{\Sigma r^3 c_s^2 \Omega} (-i\omega + \alpha \Omega \mp iK\Omega).
\ea
In practice, the modes come in pairs (one with $W_+=0$, the other
with $W_-=0$), related by $W_+=W_-^*$, $G_+=G_-^*$ and
$\omega_+=-\omega_-^*$. The real part of 
${\bf \tilde l}$ and ${\bf \tilde G}$ 
are the same for each mode in the pair, and the physical content of the
two modes is thus identical. Accordingly, we only have to consider one
set of modes.  From now on, we will assume $\tilde l_y=i \tilde l_x$ 
(a pure $W_-$ mode, with $W_+=0$).

Numerical solutions to a similar set of equations have already been
obtained for circumstellar discs in binary systems (Lubow \& Ogilvie 2000).
Compared with circumbinary discs, which will be our main application here, 
an important difference is that for circumstellar discs the eigenmodes
are a mixture of $W_+$ and $W_-$, because the driving of the disc at
the orbital frequency $\Omega_b$ of the binary cannot always be neglected
(Lubow \& Ogilvie 2000 in fact solved the equations in a frame
corotating with the binary), i.e. the assumption that the potential can
be time-averaged breaks down.  Lubow \& Ogilvie (2000) showed that this
additional effect is small for most disc parameters, but can cause 
resonances 
at specific locations in the the outer region of the disc,
leading to a growing misalignment between the circumstellar disc and the binary.

In the following, we will focus on circumbinary discs, although the
equations derived here are valid for any configuration in which the
perturbing potential can be time-averaged -- we will discuss the
behavior of circumstellar discs away from resonances in
Sec.~\ref{sec:circumstellar}.  In Sec.~\ref{sec:numsol} we present
numerical solutions for the warp equations and the inclination damping
rate. To gain more insight into the properties of the solutions for
generic external potentials and disc profiles, however, it is useful
to obtain analytical approximations to the numerical results. In the
following subsections, we discuss such an approximate treatment for
the lowest order warp mode of the disc -- which in general has the
slowest damping rate, and will thus dominate the long-term evolution
of the warp (high-order modes are also more likely to violate the
small warp condition, and are thus likely to damp even faster than
predicted in the linear theory).

\subsection{Approximate analytical solutions}
\label{sec:ana}

To obtain an approximate solution for the lowest order eigenmode of
the warp, we assume that this mode is a small correction to the flat
disc configuration.  It is also convenient to rotate the coordinates
to a frame following the twist of the disc.
Let the warp angle be $l_\theta(r)=\sin{\beta}(r)$ and the twist angle be $\phi(r)$, so that 
$l_x+il_y=l_\theta (r) \exp[i\phi(r)]$.
Define the unit vectors
\ba
\hat{e}_{\theta}&=&\cos{[\phi(r,t)]} \hat{e}_x + \sin{[\phi(r,t)]} \hat{e}_y\\
\hat{e}_{\phi}&=&-\sin{[\phi(r,t)]} \hat{e}_x + \cos{[\phi(r,t)]} \hat{e}_y.
\ea
We consider a mode with $\tilde l_y = i \tilde l_x$ and $\tilde{G}_y =
i \tilde{G}_x$, and solve for the disc profile at $t=0$ (i.e. the real part of $\tilde l,\tilde G$). 
We also define
\be
\omega = \omega_p - i \gamma,
\ee
with $\omega_p$ the precession rate of the disc, and $\gamma$ the damping rate 
of the disc inclination. 
In that frame, $l_\phi=0$ (by the definition of $\hat{e}_\phi$) 
and the eigenmode satisfies the following (real) equations
\ba
\frac{dG_{\phi}}{dr}&=& \Sigma r^3 \Omega [\omega_p-Z\Omega] l_\theta - \frac{d\phi}{dr} G_\theta\\
\frac{dG_{\theta}}{dr}&=&-\gamma\Sigma r^3 \Omega l_\theta + \frac{d\phi}{dr} G_\phi\\
l_\theta \frac{d\phi}{dr}&=&  \frac{4}{\Sigma c_s^2 r^3 \Omega} \left( [\alpha \Omega - \gamma]G_{\phi} + [\omega_p - K\Omega]G_\theta\right)\\
\frac{dl_\theta}{dr}&=&\frac{4}{\Sigma c_s^2 r^3 \Omega}\left([K\Omega-\omega_p]G_\phi + [\alpha \Omega - \gamma] G_\theta \right)
\ea
To the lowest order in the perturbation from the flat disc solution
($G_\theta=G_\phi=\phi=\gamma=0$, and $l_\theta=$constant, which we set
to unity without loss of generality as the equations are homogeneous),
we have
\ba
\frac{dG_{\phi}}{dr}&=& \Sigma r^3 \Omega [\omega_p-Z\Omega]\label{eq:Gphi}\\
\frac{d\phi}{dr}&=&  \frac{4\alpha}{\Sigma c_s^2 r^3 } G_{\phi}\label{eq:phi}\\
\frac{dl_\theta}{dr}&=&\frac{4}{\Sigma c_s^2 r^3 \Omega}[K\Omega-\omega_p]G_\phi\label{eq:lth}\\
\frac{dG_{\theta}}{dr}&=&\frac{4 \alpha G_\phi^2}{\Sigma c_s^2 r^3}-\gamma\Sigma r^3 \Omega \label{eq:Gth}.
\ea
This solution is valid as long as $| l_{\theta}(r)- 1| \ll 1$ for all
$r$, $d\phi/dr$ is small enough that the small warp condition
$|d\hatl/d\ln r|\ll 1$ 
is satisfied, and the damping rate satisfies $\gamma \ll \alpha \Omega$.

Integrating Eq.~(\ref{eq:Gphi}) [or Eq.~(\ref{eq:Gx})] over the entire
disc, and taking into account the zero-torque boundary condition, we obtain
the global precession frequency of the disc:
\be
\omega_p = \frac{\int_{r_{\rm in}}^{r_{\rm out}}d\tilde{r} \Sigma \tilde{r}^3 \Omega^2 Z}{\int_{r_{\rm in}}^{r_{\rm out}}d\tilde{r} \Sigma \tilde{r}^3 \Omega}.
\ee
To the lowest order, the disc 
precesses as a rigid-body at the frequency
$\omega_p$, and there is no damping of the warp.  
For $Z>0$, the precession will be prograde, while
for $Z<0$ the precession is retrograde (for circumbinary and circumstellar discs, we
have $Z<0$).

The first-order correction to the internal stress $\tilde \Gv$ is
obtained by integrating Eq.~(\ref{eq:Gphi}), giving
\be
G_\phi(r) = \int_{r_{\rm in}}^r d\tilde{r} \Sigma \tilde{r}^3 \Omega (\omega_p - Z\Omega)\label{eq:iGphi}.
\ee
The stress is $90^0$ out of phase with the warp, linear in the
external torque [i.e. in $Z(r)$] and, by construction, vanishes on
both boundaries.


We are now free to choose boundary conditions for ($l_\theta,\phi$), as long
as the resulting disc profile is consistent with our perturbative analysis.
A good choice is to fix the orientation of the disc at the outer boundary,
$l_\theta(r_{\rm out})=1,\phi(r_{\rm out})=0$.
We then integrate Eqs.~(\ref{eq:phi})-(\ref{eq:lth}) to find
\ba
\phi(r) &=& \int_{r_{\rm out}}^r d\tilde{r} \frac{4 \alpha}{\Sigma c_s^2 \tilde{r}^3} G_\phi, \label{eq:intphi}\\
l_\theta(r)&=& 1+ \int_{r_{\rm out}}^r d\tilde{r} \frac{4}{\Sigma c_s^2 \tilde{r}^3 \Omega} \left(\Omega K - \omega_p \right) G_\phi.
\ea
We thus have a twist of the disc proportional to the viscosity $\alpha$, and a
small warp due to the external torque.  The dimensionless warp
$\psi=|\partial \hatl/\partial (\ln{r})|$ is then easily obtained (reintroducing the scaling of the warp
with the misalignment angle of the outer edge of the disc $\beta_{\rm out}$,
neglected when we set $l_\theta=1$):
\be
|\psi| = |G_\phi| \frac{4}{\Sigma c_s^2 r^2 \Omega} \sqrt{\alpha^2 \Omega^2 + (K\Omega-\omega_p)^2}\sin{\beta_{\rm out}}.
\ee
This expression will make it easy to check whether the small warp
condition $\psi \ll 1$ is satisfied 
so that the linear theory is valid.

Up to this point, our solution for $\omega_p$ and $l_\theta$ is formally
equivalent to the results derived by Papaloizou \& Terquem (1995) in
the context of circumstellar discs. 
By including the effect of 
finite viscosity $\alpha$, however, we have also derived the twist of the
disc $\phi(r)$ and can now compute the damping rate of the mode $\gamma$. 
Integrating Eq.~(\ref{eq:Gth}), we find
\be
\gamma =\frac{ \int_{r_{\rm in}}^{r_{\rm out}}d\tilde{r} 
\frac{4 \alpha G_\phi^2}{\Sigma c_s^2 \tilde{r}^3}}
{\int_{r_{\rm in}}^{r_{\rm out}}d\tilde{r} \Sigma \tilde{r}^3 \Omega}
\label{eq:gamma},
\ee
which is effectively equivalent to what is found at first order when using the expansion
of the solution in powers of the tidal potential proposed by Lubow \& Ogilvie (2000). 

We see that $\gamma \geq 0$ for all values of $G_\phi$, i.e., the
lowest-order warp mode
is always damped when considering time-averaged axisymmetric perturbing potentials.
A simple physical interpretation of $\gamma$ can be obtained by integrating 
the numerator of Eq.~(\ref{eq:gamma})
by parts and using Eqs.~(\ref{eq:iGphi})-(\ref{eq:intphi}). We find
\be
\gamma = \frac{ \int_{r_{\rm in}}^{r_{\rm out}}d\tilde{r} \Sigma \tilde{r}^3 \Omega (Z\Omega - \omega_p) \phi }
{\int_{r_{\rm in}}^{r_{\rm out}}d\tilde{r} \Sigma \tilde{r}^3 \Omega}.
\ee
Here the denominator is the total angular momentum of the disc $L_{\rm disc}$
(up to a factor of $2\pi$). 
The numerator is the difference between the total torque applied on the disc, and the
torque required to maintain the global precession of the disc (up to
the same factor of $2\pi$). The existence of this difference is
due to the fact that the finite viscosity $\alpha$ twists the
disc by an angle $\phi$, thus creating a non-zero torque on the disc
in the plane defined by the orbital angular momentum of the binary
${\bf L}_b$ and that of the disc.

We can similarly infer the back-reaction torque applied by the 
warped disc onto the binary,
\be
{\bf T}_{\rm br} = -2\pi \int_{r_{\rm in}}^{r_{\rm out}}d\tilde{r} \Sigma \tilde{r}^3 Z\Omega^2 
({\bf \hat{e}}_z \times \hatl).
\ee
Recall that ${\bf L}_b$ is along ${\bf\hat{e}}_z$, while $\hatl$ generally deviates
from ${\bf L}_{\rm disc}$ (the total angular momentum vector of the disc).
The term in ${\bf T}_{\rm br}$ along the direction 
${\bf L}_{\rm disc}\times {\bf L}_b$ causes ${\bf L}_b$ to precess around 
${\bf L}_{\rm disc}$, while the term in the plane
defined by ${\bf L}_{\rm disc}$ and ${\bf L}_b$ can either cause the
binary to align with the disc (for prograde discs) or drive an
increase in the misalignment between the disc and the binary (for
retrograde discs). The relative importance of 
the back-reaction torque
compared to the ``direct'' damping effect of the disc inclination
depends on the ratio of the angular momenta of the disc and the binary
($L_{\rm disc}$ and $L_b$).
If $L_b\gg L_{\rm disc}$
the disc will align with the binary on a
timescale $\sim \gamma^{-1}$. For arbitrary ratio $L_{\rm disc}/L_b$, 
we must take into account changes to the angular momenta of both the disc and
the binary, and alignment occurs on the timescale
\be
t_{\rm damp} \simeq \gamma^{-1} (1+L_{\rm disc}/L_b)^{-1}.
\ee

We could, in theory, iterate further and compute the second-order
corrections to the warp, stress, and eigenfrequency. However, if the
first-order corrections are large enough that this appears warranted,
there is no guarantee that this procedure will eventually converge.
Solving the eigenvalue problem numerically is then probably a
safer option. When the first-order corrections are small, however, we
will see that the approximate formulae derived above are
accurate enough to give good estimates of the warp and twist of the
disc and of the damping timescale of its lowest-order eigenmode.

\subsection{Analytical Results for Specific disc Models}

\begin{figure}
\includegraphics[width=8.3cm]{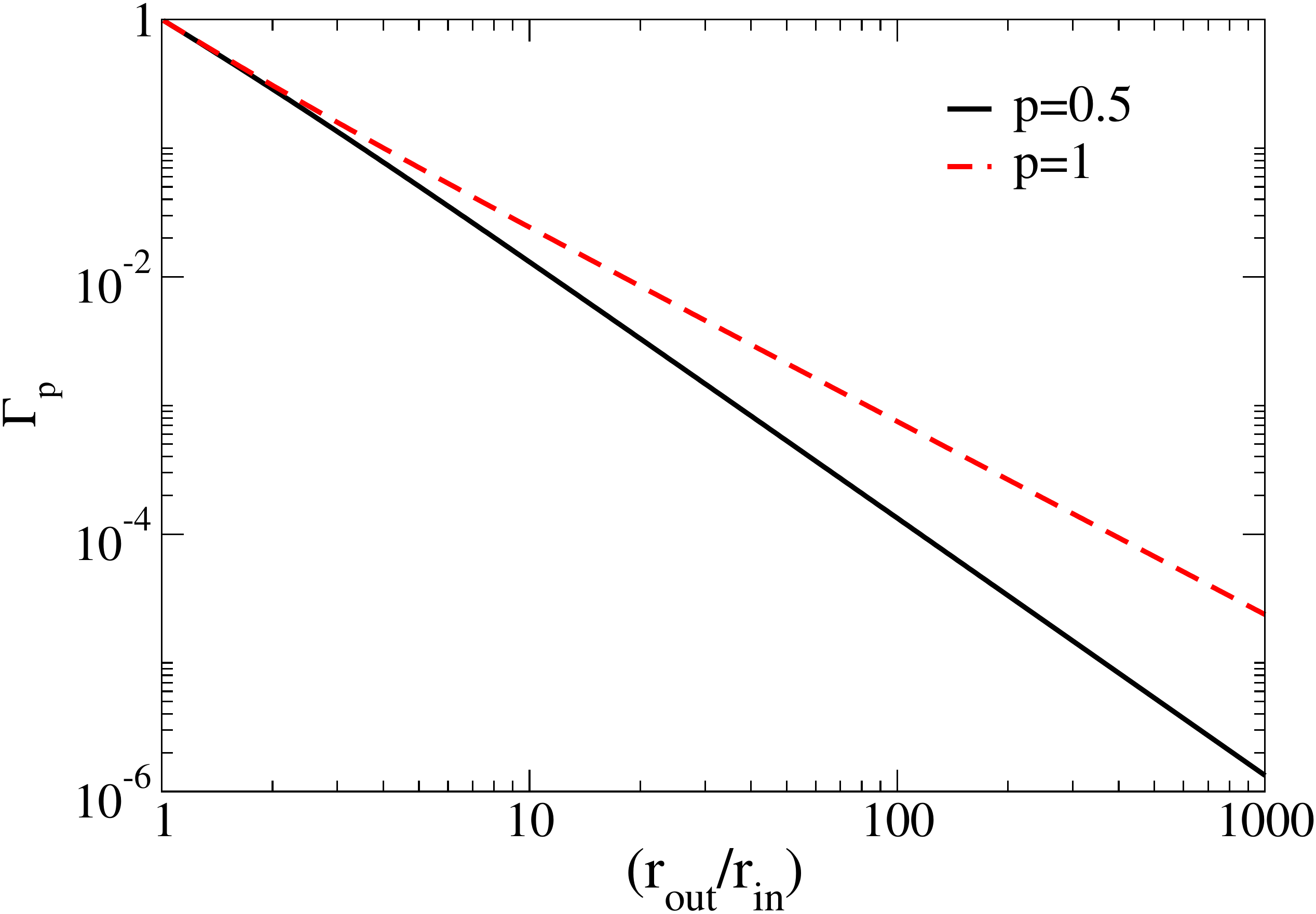}
\caption{Shape function $\Gamma_p$ determining the global precession rate of the disc
[see eq.~(\ref{eq:Gamma_p})]
for circumbinary discs with $p=0.5$ and $p=1$ ($\Gamma_p$ is independent of the sound
speed profile).}
\label{fig:GammaP}
\end{figure}

\begin{figure}
\includegraphics[width=8.3cm]{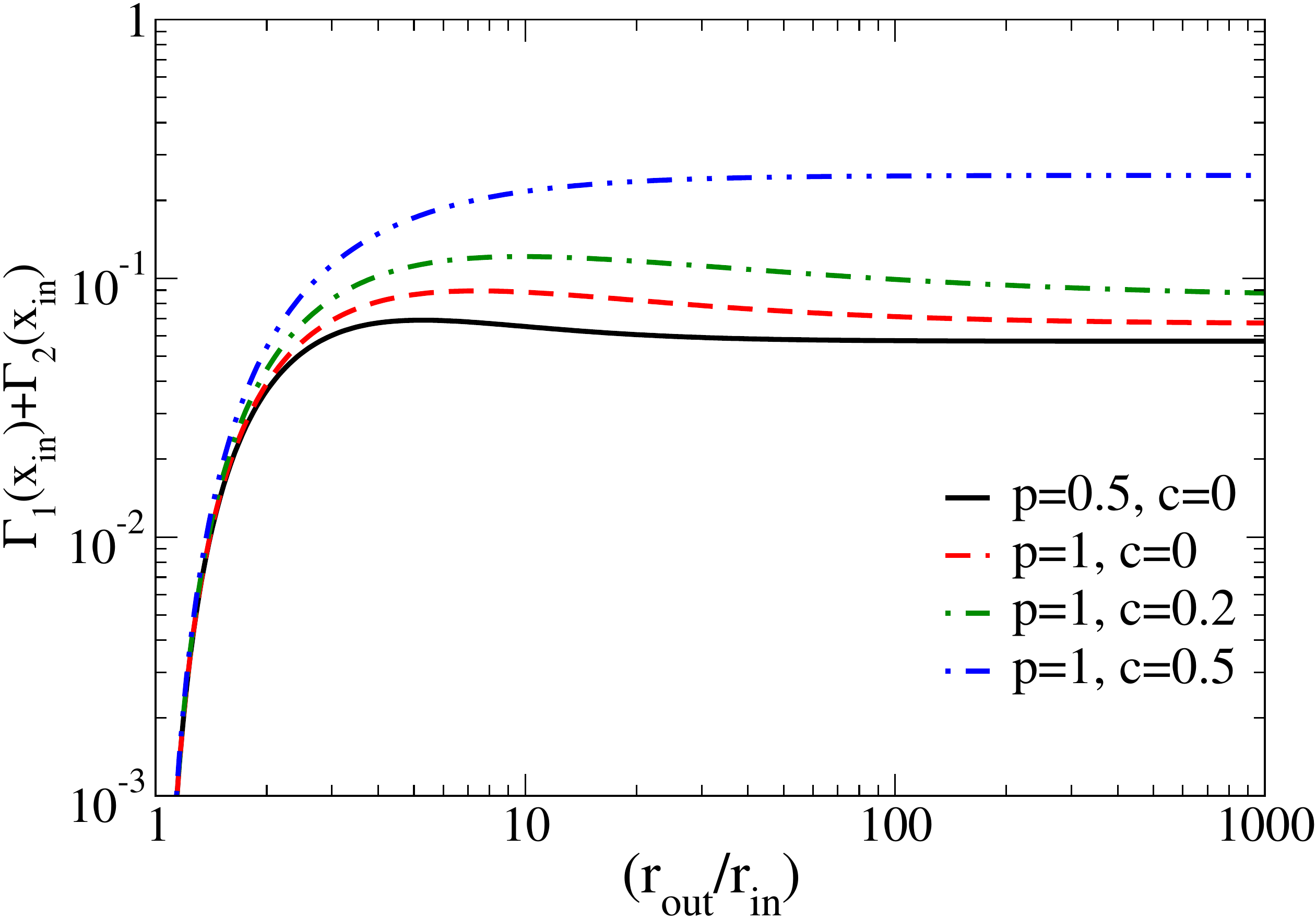}
\caption{Shape function $\Gamma_1+\Gamma_2$ determining the warp of the disc
[see eq.~(\ref{eq:Gamma_12})]
for circumbinary discs with $p=0.5$, $p=1$ and a constant sound speed, as well as for
discs with $p=1$ and different sound speed profiles $c=0.2,0.5$ (a flaring disc and a 
disc of constant thickness $\delta=H/r$). The functions are evaluated at the inner edge
of the disc, in order to provide the total warp of the disc.}
\label{fig:Gamma12}
\end{figure}

\begin{figure}
\includegraphics[width=8.3cm]{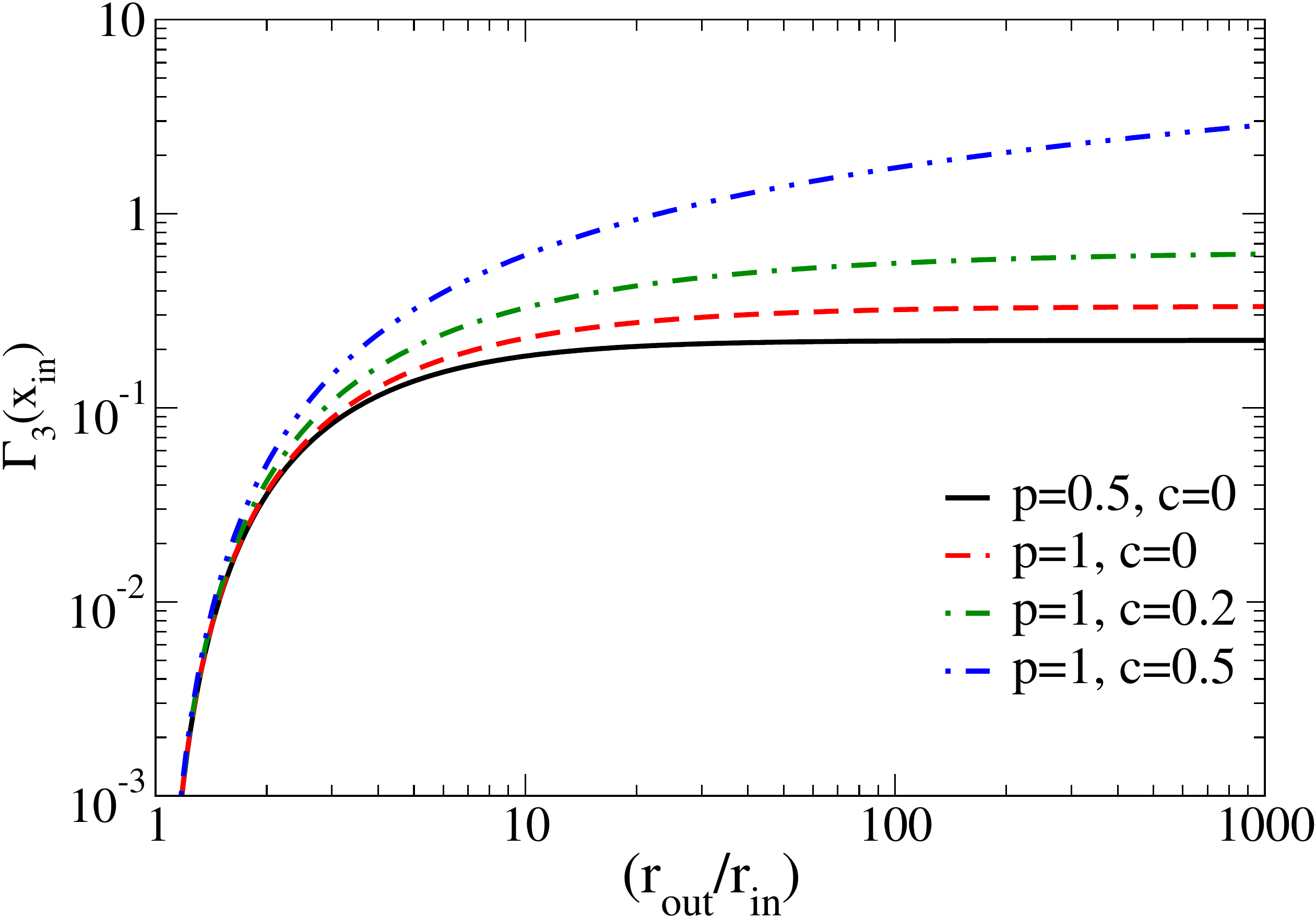}
\caption{Shape function $\Gamma_3$ determining the twist of the disc
[see eq.~(\ref{eq:Gamma_3})]
for the same discs as in Fig.~\ref{fig:Gamma12}. The functions are evaluated at the inner edge
of the disc, in order to provide the total twist of the disc}
\label{fig:Gamma3}
\end{figure}

\begin{figure}
\includegraphics[width=8.3cm]{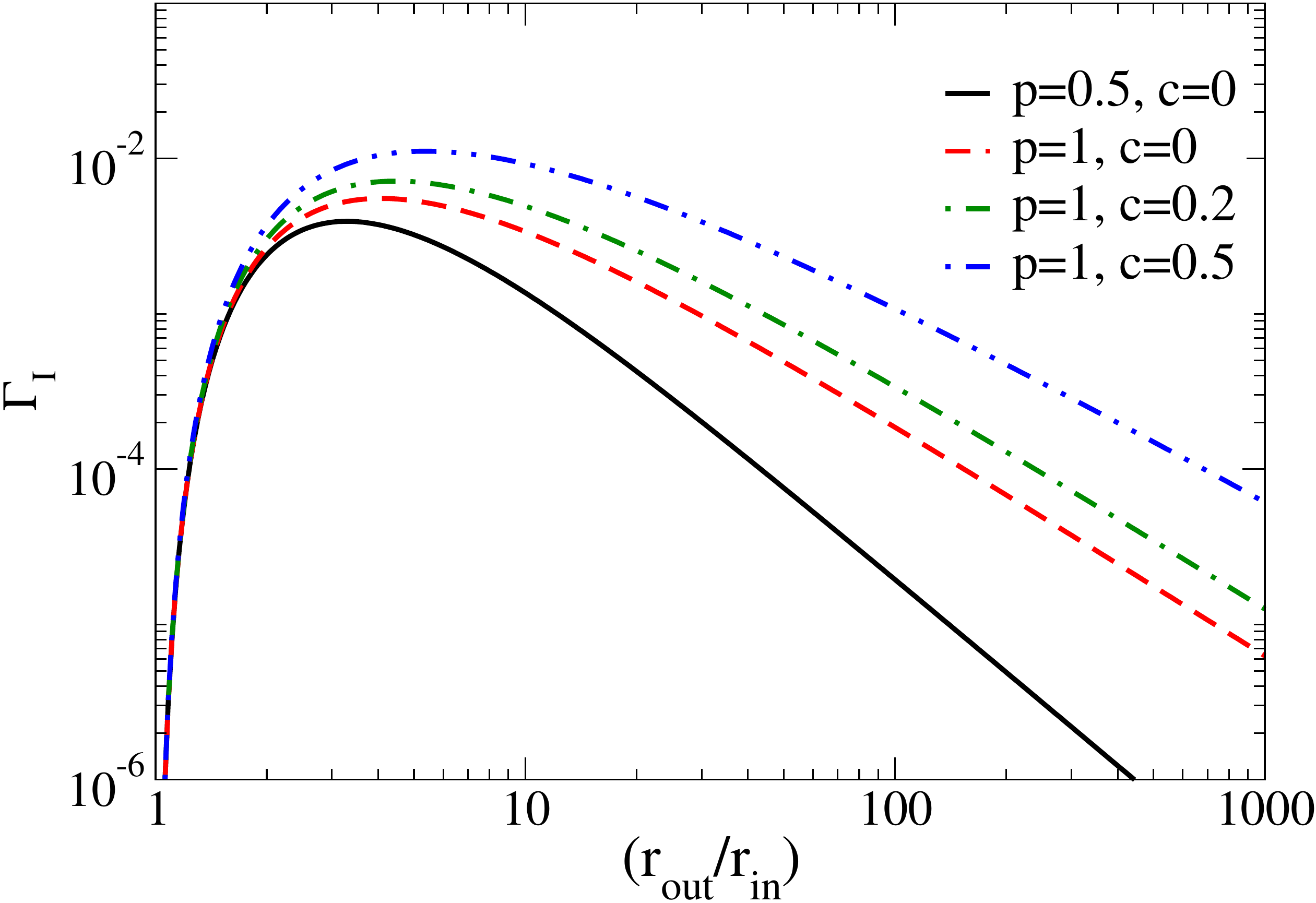}
\caption{Shape function $\Gamma_I$ determining the alignment timescale
of the disc 
[see eq.~(\ref{eq:Gamma_I})]
for the same discs as in Fig.~\ref{fig:Gamma12}.}
\label{fig:GammaI}
\end{figure}

The analytical expressions given in Sec.~\ref{sec:ana} are
significantly easier to compute than the full numerical solutions to
the eigenvalue problem. However, they still require recursive
computations of integrals over the whole disc.
Here we consider discs with the following power-law profiles:
\ba
&&\Sigma(r) = \Sigma_{\rm in} \left(\frac{r}{r_{\rm in}}\right)^{-p},
\quad c_s(r) = c_{\rm in} \left(\frac{r}{r_{\rm in}}\right)^{-c},\\
&&Z(r) = Z_{\rm in} \left(\frac{r}{r_{\rm in}}\right)^{-z}, 
\quad K(r) = K_{\rm in} \left(\frac{r}{r_{\rm in}}\right)^{-k},\\
&&\Omega(r)=\Omega_{\rm in}\left(\frac{r}{r_{\rm in}}\right)^{-3/2}.
\ea
Substituting these profiles into the relevant expressions of 
Sec.~\ref{sec:ana}, we obtain (where $x=r/r_{\rm in}$)
\ba
\omega_p &=& Z_{\rm in} \Omega_{\rm in} \Gamma_p,\label{eq:Gamma_p}\\
G_\phi(x) &=& -\Sigma_{\rm in} r_{\rm in}^4 \Omega_{\rm in}^2 Z_{\rm in} \Gamma_G(x), \\
l_\theta(x) &=& 1-\frac{4Z_{\rm in}}{\delta_{\rm in}^2} (Z_{\rm in} \Gamma_1(x) -
K_{\rm in}\Gamma_2(x)),\label{eq:Gamma_12}\\
\phi(x)&=&  \frac{4\alpha Z_{\rm in}}{\delta_{\rm in}^2} \Gamma_3(x),\label{eq:Gamma_3}\\
\frac{|\psi|}{\sin{\beta_{\rm out}}} &=& \frac{4 |Z_{\rm in}|}{\delta_{\rm in}^2}  x^{p+2c-2} \Gamma_G(x)\sqrt{\alpha^2 + (K-\frac{\omega_p}{\Omega})^2} \label{eq:plpsi},\\
\gamma &=& \frac{4 \alpha \Omega_{\rm in} Z_{\rm in}^2}{\delta_{\rm in}^2} \Gamma_I
\label{eq:Gamma_I}\ea
for the precession frequency, internal stress, warp, twist and damping
frequency of the eigenmode. In the above, 
$\delta_{\rm in}=c_{\rm in}/(r_{\rm in}\Omega_{\rm in})$ is the
thickness of the disc at its inner edge.
The shape functions $\Gamma_{1,2,3,p,G,I}$, which depend only
on the shape of the background profiles and on the relative size of the disc
$x_{\rm out}=r_{\rm out}/r_{\rm in}$,
are given in Appendix~\ref{App:ShapeFunctions}. 
\footnote{
We here choose a notation in which the dependency of the shape functions $\Gamma$ 
in $x_{\rm out}$ and in the disc profile is implicit, to simplify the equations.
}
Their signs are chosen
so that $\Gamma>0$ for circumbinary discs. For convenience, we also plot
their value for a few disc profiles important to the study of circumbinary discs
($z=k=2$, $Z=-K$, $p=0.5$ or $p=1$, and $c=0,0.2,0.5$, see discussion below),
in Figs.~\ref{fig:GammaP}-\ref{fig:GammaI}.

An approximate expression for the mode damping rate $\gamma$, based on estimating the
energy loss rate due to viscous dissipation in the disc, has previously
been proposed in the context of circumstellar discs
(using arguments generally applicable to both circumstellar and circumbinary discs)
by Bate et al.~(2000) for discs of constant thickness
($\delta=H/r=$constant):
\be
\gamma_{\rm Bate} = \frac{\alpha}{\delta^2}\frac{\omega_p^2}{\Omega(r_{\rm out})}=\frac{\alpha Z_{\rm in}^2 \Omega_{\rm in}}{\delta^2} \Gamma_p^2 x_{\rm out}^{3/2}.
\ee
Not surprisingly, our results agree with this expression 
as far as the general scaling of $\gamma$ with $\alpha$, $\delta_{\rm in}$ and 
$Z_{\rm in}$ is concerned. The actual value of $\gamma$ can
however vary by a factor of a few for moderate-sized discs ($x_{\rm out}\sim 10$), 
and by more than an order of magnitude for $x_{\rm out}\sim 100-1000$.  
The approximation used in Bate et al.~(2000) only
recovers the correct scaling of $\gamma$ with
the size of the disc in the special case $p=1,c<0.5$, and does not
find the asymptotic behavior $\gamma \rightarrow 0$ for an infinitely narrow disc 
$x_{\rm out}\rightarrow 1$ (when $\Gamma_p\rightarrow 1$, but
$\Gamma_I\rightarrow 0$).  We will show that our formula agrees 
much better with the numerical simulations of warped
discs, and with the exact solution to the eigenvalue problem.

Our results are also in agreement with the simpler model that we
derived in a previous paper in the limit of $x_{\rm out}\rightarrow
\infty$ and $\omega_p=0$ (Foucart \& Lai 2013).

\subsection{Numerical solutions: disc eigenmodes}
\label{sec:numsol}

Before applying our analytical results to real astrophysical systems, we
would like to get an estimate of their accuracy when compared to the
numerical solution of the eigenvalue problem given by
Eqs.~(\ref{eq:Gx})-(\ref{eq:ly}). 
We use the shooting method to solve the set of 4 complex
ODEs, combined with a Newton-Raphson root-solver to find the value of
the warp at the disc boundaries and the complex eigenvalue
$\omega$. In more details, we assume the boundary conditions
$\tilde l_x(r_{\rm out})=1$, $\tilde G_{x,y}(r_{\rm in})=0$ and 
$\tilde G_{x,y}(r_{\rm out})=0$, and we solve for
$\tilde l_y(r_{\rm out})$, $\tilde l_{x,y}(r_{\rm in})$ and $\omega$
using the initial guess $\tilde l_y(r_{\rm out})=i$, $\tilde
l_{x,y}(r_{\rm in}) = \tilde l_{x,y}(r_{\rm out})$,
$\omega=10^{-7}\Omega_{\rm in}$. At each Newton-Raphson iteration, we
integrate the 4 complex ODEs from both boundaries, and measure the
mismatch between the solutions at the midpoint of the disc. The
problem is then reduced to solving for the value of the 4 complex
unknowns, $\tilde l_y(r_{\rm out})$, $\tilde l_{x,y}(r_{\rm in})$ and
$\omega$, which satisfy the 4 complex matching conditions at the
midpoint of the disc
\footnote{Note however that we can always make use of the fact that
  $\tilde l_y=i \tilde l_x$ for the modes that we are interested in -
  so that in practice the number of independent variables and
  equations could be reduced by a factor of 2.
  Our code solves the full system of equations because it was
  developed in a more general framework in which the two families
  of modes $W_{\pm}$ might be distinct (see e.g. Lubow \& Ogilvie 2000).  
  }.  In general, there is an
infinite number of eigenmodes which are solutions of this problem;
which one the Newton-Raphson solver converges to depends on the chosen
initial guess for the warp on the boundaries and the eigenfrequency.
The initial choice made here is convenient for finding the solution
closest to the flat disc profile, which has the smallest warp,
internal stresses, and damping timescale rate and should thus dominate
the long term evolution of the warp. Higher-order modes can be
found using the same method if the initial guess is modified
appropriately.

We test our analytical results on a sequence of circumbinary
disc models with power-law profiles. For a circumbinary disc, we have
\ba
Z(x) &\approx& -\frac{3\eta}{4}\left(\frac{a}{r_{\rm in}}\right)^{2} x^{-2},\\
K(x) &\approx& -Z(x),
\ea
where $\eta = M_1M_2/(M_1+M_2)^2$ is the symmetric mass ratio of the
binary, and $a$ its semi-major axis.  The other parameters are set to
$p=1,c=1/2,\alpha=0.01,\delta_{\rm in}=0.1,r_{\rm in}=2a$.
For this choice of $c$, the condition for the propagation
of bending waves $\alpha<H/r$ is satisfied at all radii
($H/r=\delta_{\rm in}$ everywhere).  
For an equal mass binary ($\eta=1/4$), we then have $K_{\rm in}=-Z_{\rm in} =
0.047$. The disc precession frequency can, for $x_{\rm out} \gg 1$, be
approximated as $\omega_p=0.035 x_{\rm out}^{-3/2} \Omega_{\rm in}$. 
By comparison, the timescale for the wave to propagate across
the disc is $t_{\rm wave}\sim (0.075 \Omega_{\rm in} x_{\rm
  out}^{-3/2})^{-1}<\omega_p^{-1}$.  The disc should thus be able to
precess coherently regardless of its size.

According to Eq.~(\ref{eq:plpsi}), the dimensionless warp parameter
$\psi$ is maximum for $x \approx 0.6 x_{\rm out}$, where $\psi
\approx 0.1 \sin{\beta_{\rm out}}$.  The small warp approximation is
thus valid regardless of the value of $\beta_{\rm out}$.

\begin{figure}
\includegraphics[width=8.3cm]{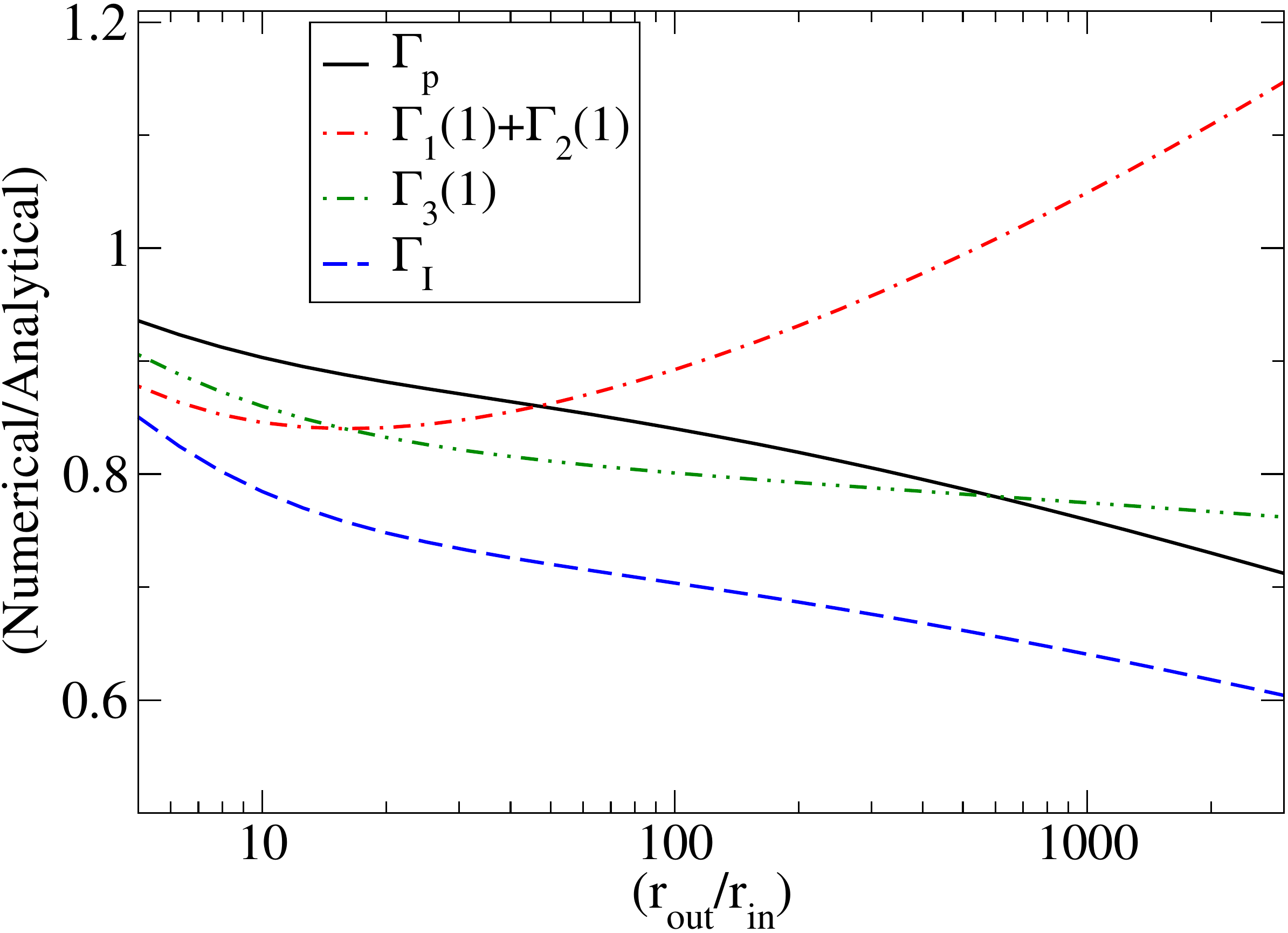}
\caption{The ratio of the numerical and analytical values of the shape
  functions $\Gamma_p$, $[\Gamma_1(1)+\Gamma_2(1)]$, $\Gamma_3(1)$ and
  $\Gamma_I$ for a misaligned circumbinary disc.  The disc is assumed
  to follow a power-law density profile with $p=1$, is of constant dimensionless
  thickness $\delta=H/r=0.1$, and has a viscosity parameter of $\alpha=0.01$.  The
  outer edge of the disc is varied between $r_{\rm out}=5r_{\rm in}$
  and $r_{\rm out}=3000r_{\rm in}$. The central binary has equal mass
  components, and a semi-major axis $a=0.5r_{\rm in}$.}
\label{fig:NumVsAna}
\end{figure}

To compare the analytical and numerical solutions, we extract from the
numerical results the shape functions $\Gamma_p,\Gamma_I,\Gamma_3(1)$,
and $[\Gamma_1(1)+\Gamma_2(1)]$ (we cannot extract separately
$\Gamma_1$ and $\Gamma_2$, as they both contribute to the warp of the
disc), and divide them by the analytical values obtained in
Appendix~\ref{App:ShapeFunctions}, and plotted in Figs.~\ref{fig:GammaP}-\ref{fig:GammaI}. 
The results are shown in
Fig.~\ref{fig:NumVsAna}. The first-order analytical approximation
recovers the disc profile and the precession timescale to $\sim
10\%-20\%$ accuracy over a wide range of disc sizes ($x_{\rm out}\sim
5-1000$). The damping timescale is recovered to a similar accuracy for
small discs ($x_{\rm out} \lesssim 10$), and within $\sim 40\%$ for
larger discs.

\begin{figure}
\includegraphics[width=8.3cm]{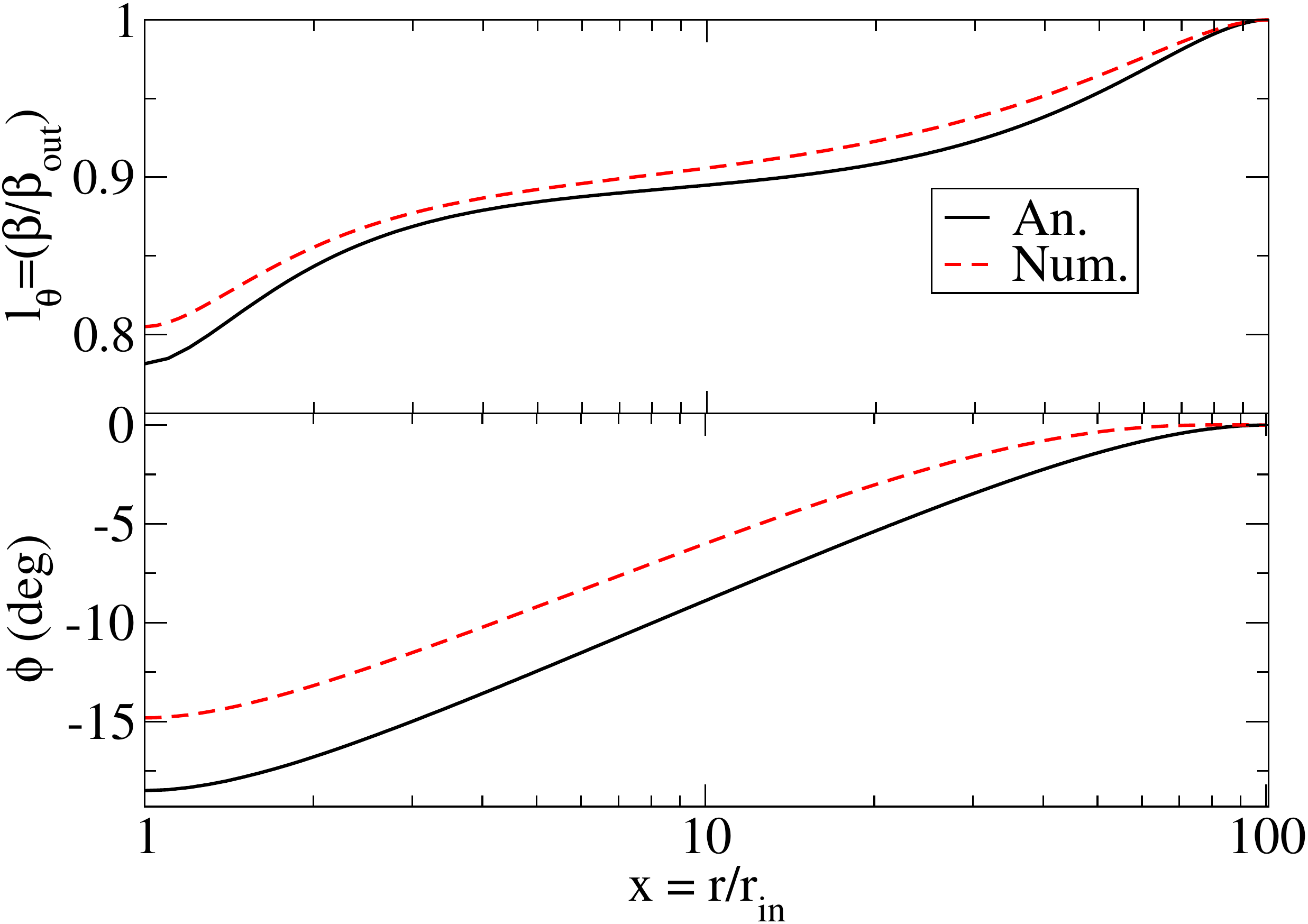}
\caption{Inclination (top) and twist (bottom) profiles for a circumbinary disc with
  $\alpha=0.01$, $\delta=0.1$, $p=1$, $c=0.5$, $r_{\rm out}=100r_{\rm
    in}$, $\eta=0.25$ and $r_{\rm in}=2a$.  The solid black lines show
  our approximate analytical solution, and the dashed red lines the
  exact numerical solution.}
\label{fig:Profile}
\end{figure}

The warp and twist profiles of the disc as a function of radius are
shown in Fig.~\ref{fig:Profile} for a disc with $x_{\rm out}=100$.  We
see that the general shape of the profile is also recovered with good
accuracy. 
The disc is warped in two distinct regions, due to the different dependence of
the shape functions $\Gamma_1$ and $\Gamma_2$ on the radius of the 
disc: at small radii,
where the non-Keplerian contributions to the warp ($\propto \Gamma_2$)
are maximum, and at larger radii, where the contributions to the warp
due to the external torque ($\propto \Gamma_1$) dominate. The twist,
on the other hand, has a single component due to the finite viscosity
of the disc ($\propto \Gamma_3$).

\begin{figure}
\includegraphics[width=8.3cm]{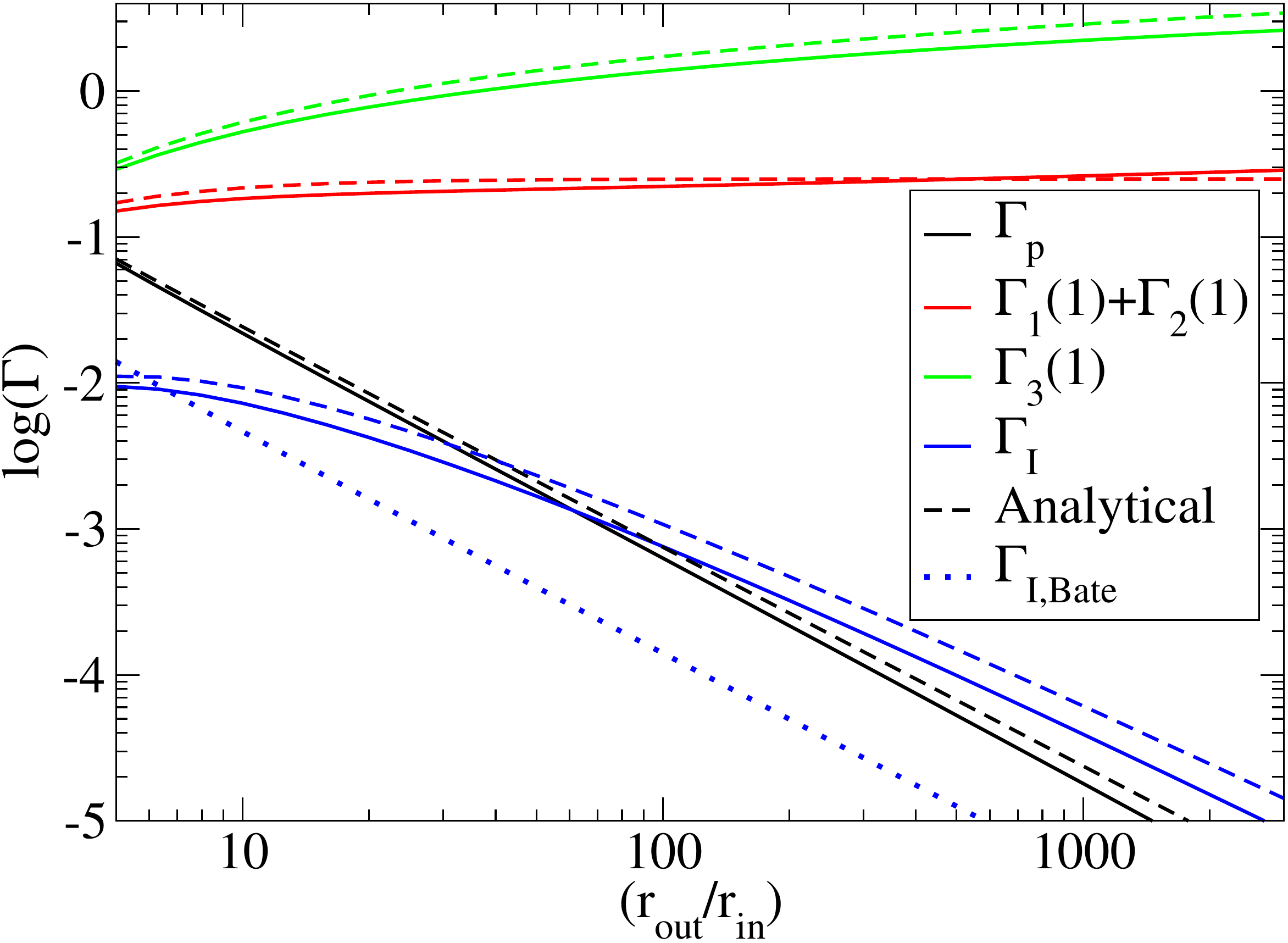}
\caption{Numerical (solid lines) and analytical (dashed lines) values
  of the logarithm of the shape functions $\Gamma_p$,
  $[\Gamma_1(1)+\Gamma_2(1)]$, $\Gamma_3(1)$ and $\Gamma_I$ for the
  same configuration as in Fig.~\ref{fig:NumVsAna}. The dotted line
  shows the damping coefficient $\Gamma_I$ corresponding to the
  approximate formula derived in Bate et al.~(2000).}
\label{fig:Shape}
\end{figure}

Finally, the values of the shape functions are plotted on
Fig.~\ref{fig:Shape} (see also Sec.~\ref{sec:std} for more
discussion).  Figure~\ref{fig:Shape} also shows the damping rate that would
be expected from the approximate formula derived in Bate et
al.~(2000). 
While the approximate formula performs relatively well for
$x_{\rm out}\sim 5-10$, it becomes increasingly unreliable for larger
discs (and for different choices of the parameters $p$ and $c$, the
error in the approximate formula
of Bate et al. can be even larger).

\section{Application to circumbinary discs}
\label{sec:circum}

\subsection{Fiducial disc Model}
\label{sec:std}

We can now use the results of the previous section to study the
properties of misaligned circumbinary discs.
For concreteness, we first consider discs with a constant thickness $H/r$
and a density profile which approximately follows a power-law with $p=1$. 
Because the shape functions 
[$\Gamma$'s in Eqs.~(\ref{eq:Gamma_p})-(\ref{eq:Gamma_I})]
depend only on $r_{\rm out}/r_{\rm in}$
and on the assumed disc profile, the other parameters
can be varied at will without changing the value of the $\Gamma$
functions --- as long as the small warp approximation remains
valid. Here we consider prograde discs with $r_{\rm out}\gg r_{\rm in}$, 
and will discuss modifications for retrograde discs in Sec.~\ref{sec:retro}.

The global precession period of the disc is
\be
P_p = 20200 \left(\frac{r_{\rm out}}{100 {\rm AU}}\right)^{3/2} \frac{1}{4\eta} \left(\frac{r_{\rm in}}{2a}\right)^{2} \left(\frac{2M_\odot}{M}\right)^{1/2} {\rm yrs}
\ee
where $M=M_1+M_2$ is the mass of the binary. The travel time of bending waves across the disc is
\be
t_{\rm wave} = 1500  \left(\frac{r_{\rm out}}{100 {\rm AU}}\right)^{3/2} \frac{0.1}{\delta_{\rm in}}  \left(\frac{2M_\odot}{M}\right)^{1/2} {\rm yrs},
\ee
which is shorter than $P_p$ for a reasonable disc thickness,
justifying global disc precession.
For comparison, the viscous timescale in the disc $t_{\nu}=r^2/(\alpha_0 H^2 \Omega)$ is
\be
t_\nu = 1.1\times 10^6\left(\frac{r}{100 {\rm AU}}\right)^{\!3/2}\!
\left(\frac{2M_\odot}{M}\right)^{\!\! 1/2} \frac{0.01}{\alpha_0}
\left(\frac{0.1}{\delta_{\rm in}}\right)^2 {\rm yrs},
\ee
which is significantly longer for our standard disc parameters. 
Note that we explicitly differentiate between the viscosity parameter $\alpha_0$
responsible for the viscous evolution of the disc density profile
and the viscosity parameter $\alpha$ responsible for the damping of disc warp.

The relative change in the warp across the whole disc, 
$\delta l_\theta=|1-l_{\theta}(1)|$, is 
\ba
\delta l_\theta &=& 0.88 \left[\Gamma_1(1)+\Gamma_2(1)
\right] (4\eta)^2 \left(\frac{2a}{r_{\rm in}}\right)^{4} \left(\frac{0.1}{\delta_{\rm in}}\right)^2
\nonumber\\
&\approx & 0.2 (4\eta)^2 \left(\frac{2a}{r_{\rm in}}\right)^{4} \left(\frac{0.1}{\delta_{\rm in}}\right)^2,
\ea
where in the second line we have used the fact 
that $\Gamma_1(1)+\Gamma_2(1)$ is nearly independent of $x_{\rm out}$ (see Fig.~\ref{fig:Shape}).
The strong dependence of the warp on the location of the inner radius
of the disc is noteworthy. If the inner edge of the disc was closer to
the binary, the small warp approximation would break down.

The twist across the disc is given by
\be
\phi(1) = -24^\circ \frac{\alpha}{0.01} (4\eta)  \left(\frac{2a}{r_{\rm in}}\right)^{2} \left(\frac{0.1}{\delta_{\rm in}}\right)^2 \frac{\Gamma_3(1)}{2}.
\ee
In contrast with the disc warp, the twist has a non-negligible
dependence on the size of the disc, since $\Gamma_3(1)$ varies by
about an order of magnitude between $x_{\rm out}=5$ and $x_{\rm out}=3000$ 
(see Fig.~\ref{fig:Shape}).  For large discs ($x_{\rm out}\sim 100-3000$), 
we have $\Gamma_3(1)\sim 2$, hence the scaling
chosen in our twist expression given above. 
However, narrower discs can have much smaller twists, 
of order of a few degrees.

For the disc profile considered in this section, Eq.~(\ref{eq:plpsi})
shows that the maximum of the dimensionless warp $\psi_{\rm max}$ is
largely independent of the location of the outer edge of the disc, and
is given by
\be
\psi_{\rm max} = 0.1 (4\eta)^2  \left(\frac{2a}{r_{\rm in}}\right)^{4} \left(\frac{0.1}{\delta_{\rm in}}\right)^2 \sin{\beta_{\rm out}}
\ee
if $\alpha \lesssim Z(1.4)$ (i.e. if we can neglect $\alpha$ in Eq.~[\ref{eq:plpsi}]).
The condition $|d\ln{\hatl}/dr|<1$ is thus easily satisfied. The stronger constraint $|d\ln{\hatl}/dr|<H/r$
suggested by the numerical work of Sorathia et al. 2013a is marginally satisfied for our fiducial disc
parameters. For thinner discs, nonlinear effects might thus cause the damping of the inclination to be
faster than what the linear theory predicts.

Finally, we can compute the alignment timescale of the disc and the binary:
\ba
t_{\rm damp}&=& \gamma^{-1} \left[1+\frac{2M_{\rm disc}}{3M} \left(\frac{a}{r_{\rm out}}\right)^{1/2} 
\right]^{-1},\\
\gamma^{-1}&=&13000 \frac{1}{x_{\rm out}^{3/2} \Gamma_I} \left(\frac{\delta_{\rm in}}{0.1}\right)^{2} \left(\frac{0.01}{\alpha}\right) \left(\frac{2M_\odot}{M_b}\right)^{1/2}\nonumber \\
 &&\times  \left(\frac{r_{\rm out}}{100 {\rm AU}}\right)^{3/2} (4\eta)^{-2}  \left(\frac{r_{\rm in}}{2a}\right)^{4} {\rm yrs}.
\ea
This means that, for $r_{\rm out}\gtrsim 100{\rm AU}$, a circumbinary
disc of constant thickness $\delta=0.1$ and viscosity $\alpha=0.01$
will align on a timescale comparable to its global precession timescale, and
much shorter than its expected lifetime. This can easily be understood
from the fact that most of the disc angular momentum is at large
radii, while most of the torque exerted by the binary on the disc is
applied at small radii. The torque is perpendicular to the local
direction of the disc angular momentum, which is
rotated by an angle $\sim \phi(1)$ with respect to the total angular
momentum of the disc. A component of order $(\sin{\phi}) T$ of the
total torque $T$ will thus attempt to align the angular momentum of
the disc, while the precession is due to a torque of order
$(\cos{\phi}) T$. For our nominal value of $\phi(1)=24^\circ$ at the
inner edge of the disc, about $0.4T$ goes towards aligning the disc
instead of driving precession.

From the above results we see that, for the fiducial disc profile considered here, an
inclined circumbinary disc can survive for a significant fraction of
its viscous evolution timescale only if one of the following conditions 
is met:
\begin{itemize}
\item A low effective viscosity $\alpha \ll \alpha_0$. This appears
  unlikely; if anything, the growth of parametric instabilities could
  cause $\alpha \gg \alpha_0$ for misalignments greater than a few
  degrees (see Sec.~\ref{sec:param}).
\item A thick disc $\delta \gtrsim 0.3$. This is also unlikely for
  protoplanetary discs.
\item A small symmetric mass ratio $\eta \ll 1/4$. This should be rare
  as $\eta$ varies slowly with the mass ratio $q=M_2/M_1$ of the
  binary for $q\sim 1$
  (to gain a factor of $3$ in $\eta$, one needs a binary with a
  mass ratio of 1:10)
\item A large binary eccentricity. This would lead to disc truncation
  at $r_{\rm in}\gtrsim 4a$ (Artymowicz \& Lubow 1994). Given the
  strong dependance of the damping timescale in the ratio $r_{\rm
    in}/a$, this would modify the damping timescale by about
    an order of magnitude.
\end{itemize}

\subsection{Dependence on the disc profile}
\label{sec:diffprof}

The power-law surface density profile $\Sigma \propto r^{-p}$ with $p=1$ 
considered in the previous subsection provides only an approximate 
description of protoplanetary discs. Other density profiles are possible
(e.g., Williams \& Cieza 2011).
There is also uncertainty on the temperature
profile of the discs. Observations suggest that protoplanetary discs
have a 'flaring' profile, where $H/r$ grows with radius. This corresponds to
a sound speed profile $c_s(r)\propto r^{-c}$, with $0 \leq c \leq 0.5$
(with preferred value $c\sim 0.2$, or $H\propto r^{1.3}$). 
Here we examine how different disc profile parameters $(p,c)$
affect the evoluton of disc warp and misalignment.

\begin{figure}
\includegraphics[width=8.3cm]{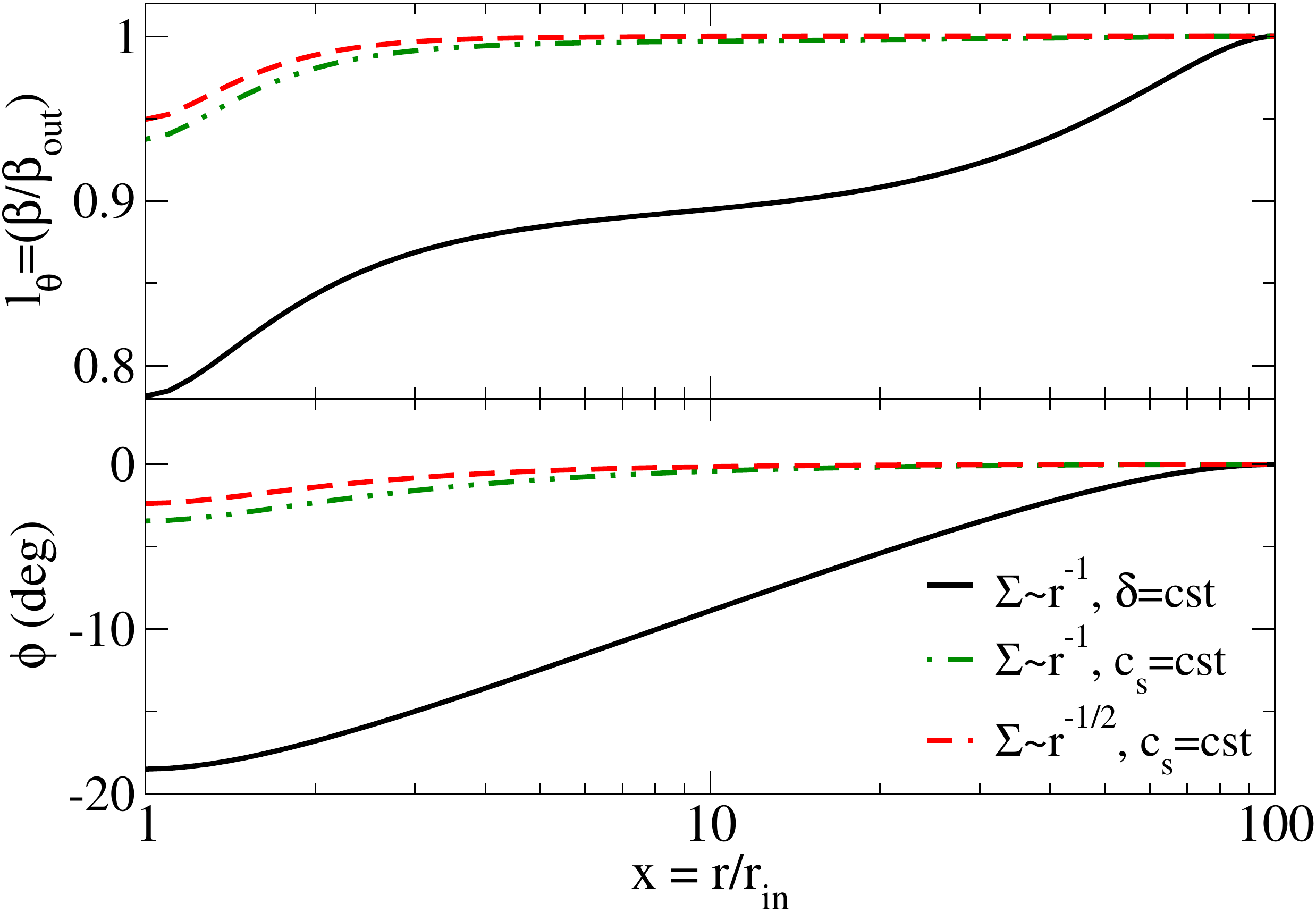}
\caption{Warp (top) and twist (bottom) profiles for 3 different
  choices of power-law profiles for discs which otherwise have
  the same parameters as in Fig.~\ref{fig:Profile}. Two of
  the profiles have $p=1$ and either a constant sound speed ($c=0$) or
  a constant $H/r$ ($c=0.5$), while the third has $p=0.5$ and a
  constant sound speed.}
\label{fig:MultProf}
\end{figure}

\begin{figure}
\includegraphics[width=8.3cm]{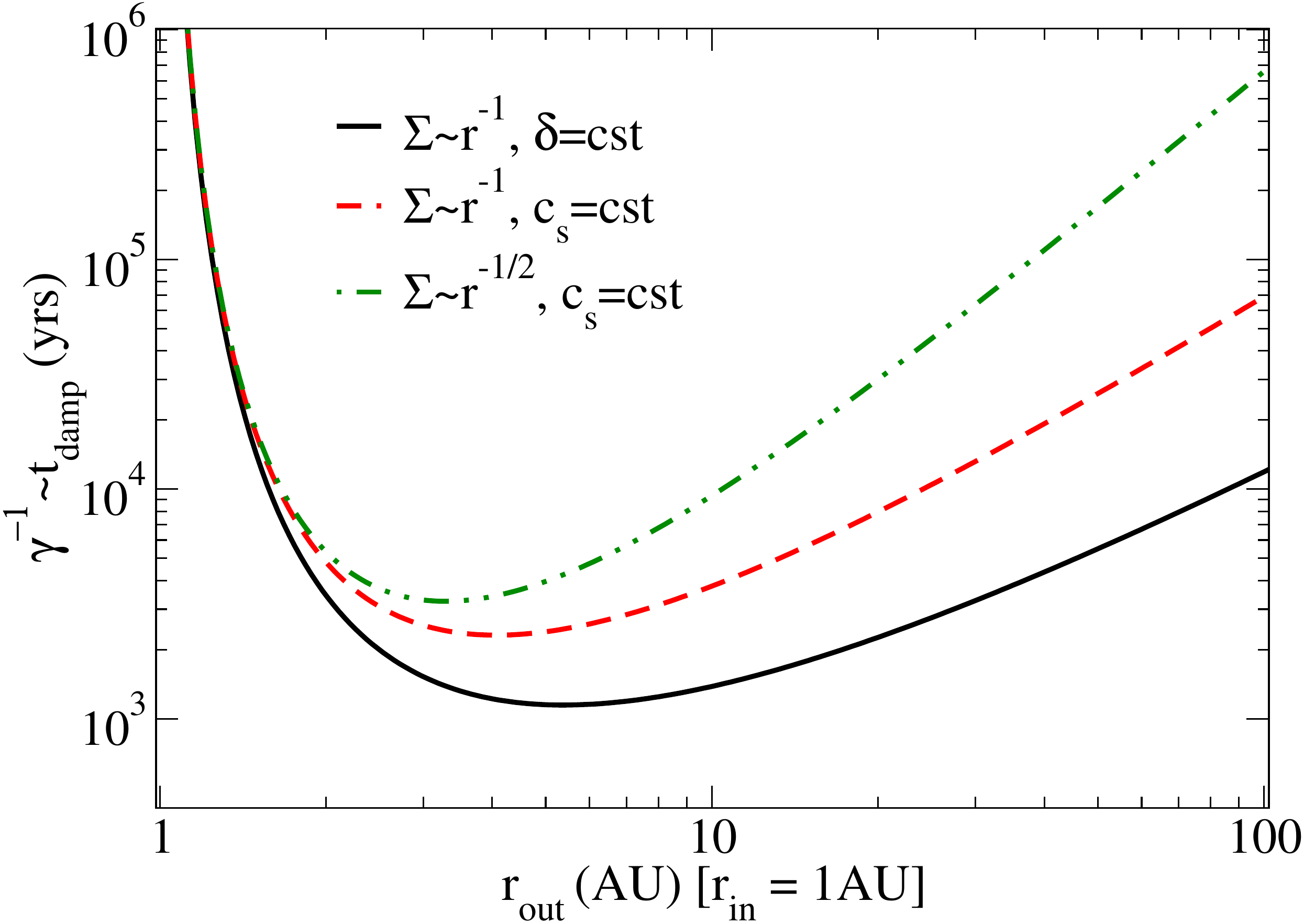}
\caption{disc inclination damping timescale as a function of the
  radial aspect ratio of the disc $x_{\rm out}=r_{\rm out}/r_{\rm in}$
  for the three choices of power-law profiles considered in
  Fig.~\ref{fig:MultProf}. The other parameters are $\alpha=0.01$,
  $\delta_{\rm in}=0.1$, $\eta=0.25$, $M=2M_\odot$ and $r_{\rm in}=2a=1{\rm AU}$.}
\label{fig:MultDamp}
\end{figure}

Qualitatively, a flaring disc has large $\delta=H/r$ for $r \gg r_{\rm in}$, 
thus decreasing the warp and twist of the disc at large
radii. A flaring disc is thus nearly flat at large radii, while its
behavior is mostly unmodified for $r\sim r_{\rm in}$.
Figure~\ref{fig:MultProf} shows an example of this behavior for the
relatively extreme case $c=0$ (constant sound speed, or $H\propto
r^{3/2}$). Because the disc is nearly flat for $r\gtrsim 5r_{\rm in}$,
the total twist and warp are about a factor of 4 smaller than in the
$c=0.5$ case (constant $\delta$).  This is reflected in the inclination damping
timescale, shown in Fig.~\ref{fig:MultDamp}: as the
damping is due to the misalignment between the outer and inner edges
of the disc, a flaring disc can remain inclined for a significantly
longer time than a disc of constant $\delta$. For $x_{\rm out} \sim 100$,
going from $c=0.5$ to $c=0$ increases the damping timescale by a factor
of $\sim 6$. We expect most protoplanetary discs to lie in
between these two limits. 

\begin{figure}
\includegraphics[width=8.3cm]{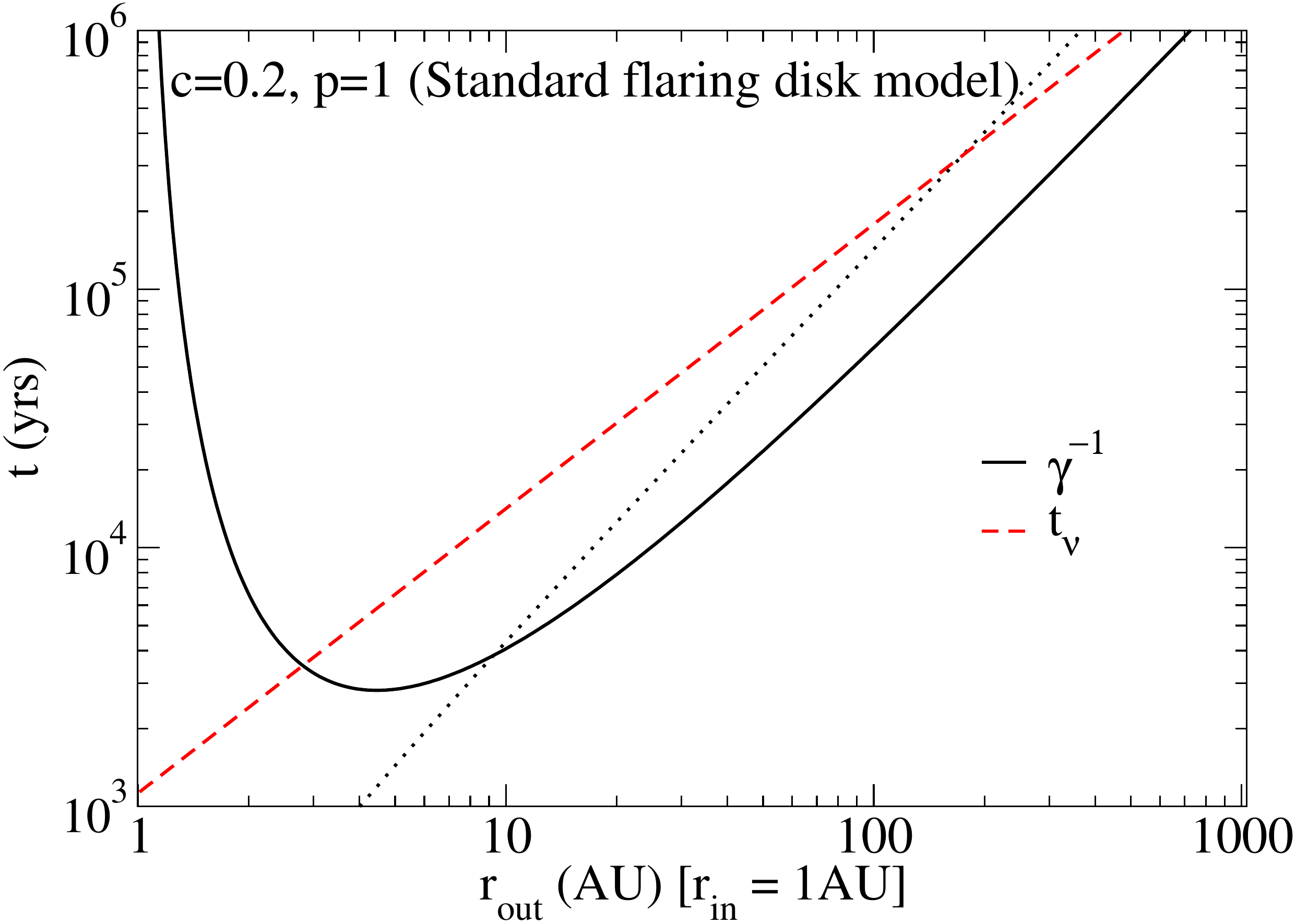}
\caption{Timescales for the damping of the inclination (solid black
  line) and the viscous evolution (dashed red line) of a typical
  flaring disc profile ($c=0.2$, $p=1$).  The other parameters of the
  system are $\alpha=\alpha_0=0.01$, $\delta_{\rm in}=0.1$,
  $\eta=0.25$, $M=2M_\odot$ and $r_{\rm in}=2a=1{\rm AU}$.  The dotted
  black line show the result based on the approximate formula of Bate et
  al.~(2000).}
\label{fig:C2}
\end{figure}

The disc flaring effect is illustrated in Fig.~\ref{fig:C2} for a 
disc model with $c=0.2$. We see that 
for our standard choice of parameters, the damping timescale for the
inclination is $\sim 20\%-30\%$ of the viscous evolution timescale
$t_\nu$ 
for $x_{\rm out} \sim 10-100$. There is thus a class of acceptable disc
parameters for which $t_{\rm damp} \sim t_\nu$. Rapid damping of the
inclination does however remain the most likely outcome.  Indeed,
$\delta_{\rm in}\sim 0.1$ is probably optimistic for such a flaring
disc, and $t_{\rm damp}/t_\nu \propto \delta_{\rm in}^4$.
Additionally, $\alpha_0 \ll 0.01$ is quite likely in the dead zones, but
the same is probably not true for $\alpha$ (see
Sec.~\ref{sec:param}). A combination of these effects is likely to
lead to the observed $t_\nu \sim 10^{6-7} {\rm yrs}$ for $r_{\rm
  out}\sim 100 {\rm AU}$ without causing as much of a modification in
$t_{\rm damp}$ -- thus leading to $t_{\rm damp} \sim (0.005-0.05)
t_\nu$ for our standard parameters. Conversely, this means that the
observation of a significant inclination in a circumbinary disc can
provide significant constraints on the age and parameters of the
system.

The density profile of the disc has a smaller effect on its warp and
twist profiles, as both the torque applied on an annulus of the disc
and the angular momentum of that annulus are proportional to its
surface density. This is shown in Fig.~\ref{fig:MultProf} for a disc
with $p=0.5$, $c=0$. The effect on the inclination damping timescale
is much more significant. Indeed, a shallower disc profile
($p<1$) has a larger fraction of its angular momentum in the outer
region of the disc. Thus, it is harder for the torques applied on the
smaller amount of material available at small radii to affect the global
evolution of the disc. Accordingly, both the precession timescale and
the damping timescale are larger (see Fig.~\ref{fig:MultDamp}). For
$x_{\rm out} \gg 1$ and $p+c<1.5$, these scale as $\omega_p \propto
\gamma \propto x_{\rm out}^{p-5/2}$.  A shallow disc profile, although
less likely than a flaring disc, thus provides another pathway to
increase the timescale over which the disc's inclination is damped.

\subsection{Retrograde circumbinary discs}
\label{sec:retro}

Most of the results discussed in the previous sections can be directly
applied to retrograde circumbinary discs, as long as the misalignment
between the orbital angular momenta of the disc and the binary remains
small. There are however a few important distinctions when
attempting to determine the long term evolution of retrograde discs.
The first is that, because there are no Lindblad resonances in these
discs, we expect the inner edge of the disc to be closer to the
binary. This will result in shorter precession ($\propto r_{\rm
in}^2$) and damping ($\propto r_{\rm in}^4$) timescales, larger
twists ($\propto r_{\rm in}^{-2}$) and warps ($\propto r_{\rm in}^{-4}$), and 
faster growth of parametric instabilities. A low mass retrograde disc
will thus rapidly reach a nearly counter-aligned configuration.

The second difference is that the back-reaction torque on the binary
has the effect of increasing the misalignment of the disc with
respect to the counter-aligned equilibrium configuration. This means
that if the angular momentum of the disc is larger than the angular
momentum of the binary, the misalignment angle is driven towards
$90^\circ$ and might eventually switch to a prograde
configuration. The details of this process have been studied under
fairly general assumptions by King et al.~(2005). 
They find that the disc will eventually become prograde 
if $\cos \theta > -L_{\rm disc}/(2L_b)$.
\section{Misaligned Circumstellar discs}
\label{sec:circumstellar}

The properties of misaligned circumbinary discs can, in the framework
presented here, easily be contrasted with those of circumstellar discs
in binary systems, as long as we neglect potential resonances between
eigenmodes of the disc and the time-dependent, $m=2$ component of the
perturbing potential (see the numerical results of Lubow \& Ogilvie
2000).  

The time-averaged potential produced by an external binary companion 
on a misaligned circumstellar disc tends to make the disc precess,
with the corresponding functions $Z(r)$ and $K(r)$ 
[see Eqs.~(\ref{eq:K(r)})-(\ref{eq:Z(r)})] given by 
\be
Z(r) = -K(r) \approx -\frac{3q}{4} \left(\frac{r}{a}\right)^3,
\ee
where $q=M_2/M_1$, 
$M_1=M_\star$ is the mass of the central star and
$M_2$ is the companion mass. 
While the warping of a circumbinary disc is mostly due to torques
applied at the inner edge of the disc, the opposite is true
for a circumstellar disc. In fact, as soon as the radial
aspect ratio of the disc is large enough ($x_{\rm out} \gtrsim 10$),
the location of the disc inner edge becomes largely irrelevant. The
global precession frequency of the disc is then 
\be
|\omega_p| = \frac{3}{8}\frac{q}{(1+q)^{1/2}}
\left(\frac{5-2p}{4-p}\right) \left(\frac{r_{\rm out}}{a}\right)^{3/2}\Omega_b.
\ee
For $r_{\rm out}\sim a/3$ (typical of the truncation radius of a circumprimary disc
due to the influence of the secondary, see Artymowicz \& Lubow 1994) and $q\sim 1$, 
the disc precesses on a timescale a few times larger than the
binary orbital period. 
The disc warp, twist and damping timescale can be computed using the expressions
given in Section 2.2. For $p=1$ and $c=0.5$ (constant $\delta$), we find
\ba
\delta l_\theta  &\approx& -0.01 \left(\frac{0.1}{\delta}\right)^2 q^2 \left(\frac{3r_{\rm out}}{a}\right)^6, \\
\phi(1) &\approx& 0.01 \left(\frac{\alpha}{0.01}\right)  \left(\frac{0.1}{\delta}\right)^2 q \left(\frac{3r_{\rm out}}{a}\right)^3, \\
\frac{\gamma}{\Omega_b} &\approx& 1.5\times 10^{-4} \left(\frac{\alpha}{0.01}\right)  \left(\frac{0.1}{\delta}\right)^2 \! \frac{q^2}{(1+q)^{1/2}} \left(\frac{3r_{\rm out}}{a}\right)^{\!9/2}.
\label{eq:gammacircumstellar}
\ea
We see that circumprimary discs generally 
have very small warps ($\lesssim 1\%$). Circumsecondary discs have higher mass ratios $q>1$,
but are also truncated at a smaller radius (Artymowicz \& Lubow 1994) and will thus avoid significant
warping.
In contrast to 
circumbinary discs, the misalignment of a circumstellar disc also damps 
on a timescale much longer than the precession timescale or even the viscous timescale.
It is thus much more natural to expect misaligned
circumstellar discs than misaligned circumbinary discs. 
Furthermore, in circumstellar discs, resonances can drive a growth
of the disc inclination for some specific choices of the ratio $r_{\rm out}/a$ 
(see Lubow \& Ogilvie 2000).

Our analytical result for $\gamma$ matches the 
numerical solution of Lubow \& Ogilvie (2000) to high accuracy
(if we adopt the same disc parameters: $p=0.5$, $c=0.5$, for which the numerical
factor in Eq.~(\ref{eq:gammacircumstellar}) is 
$10^{-4}$ instead of $1.5\times 10^{-4}$).
This is not surprising given that we are starting from the same system of
equations and that the assumption of linear warps is more
accurate for circumstellar discs than circumbinary discs.\footnote{Note, however,
  that we define the effective scaleheight $H$ through $P=\Sigma
  \Omega_z^2 H^2$, as in Lubow et al. (2002), while it is defined as
  $P=\Sigma \Omega_z^2 H^2 / (2n+3)$ in Lubow \& Ogilvie (2000), where
  $n$ is the polytropic index of the fluid.  Hence, for the $n=3/2$
  polytrope used in Lubow \& Ogilvie (2000), $H$ should be modified by
  a factor of $\sqrt{6}$ 
when comparing their results to ours.}
Both the results of Lubow \& Ogilvie (2000) and those presented here show damping
timescales significantly shorter than what would be found using the approximate formula 
of Bate et al. (2000), by up to an order of magnitude in the range of parameters considered here.

\section{Parametric instabilities}
\label{sec:param}

\subsection{Conditions for the growth of parametric instabilities}

The bending waves by which 
warps propagate in discs with $H/r\go\alpha$ are
associated with strongly shearing epicyclic motions, whose velocity is
proportional to the distance above the mid-plane of the disc
(Paploizou \& Pringle 1983). When those velocities become close to the
sound speed in the disc at some characteristic height $z\sim H$, 
Gammie et al.~(2000) have shown that the
internal shearing motions induced by the bending waves are locally
unstable to parametric instabilities on a timescale of order 
the orbital period of the disc. Even when those velocities are
subsonic, the instabilities grow on a timescale $H/v'$ (where $v'$
is the velocity of the shearing motions
for $z\sim H$) as long as this growth rate
is faster than the viscous damping timescale $H^2/\nu$.
This translates to the condition
\be
|\psi| \geq \max{\left(A_1\left|q-\frac{3}{2}\right|
\alpha,A_2\alpha^2\right)}, \label{eq:param}
\ee
where $q=-d(\ln{\Omega})/d\ln{r}$ and $(A_1,A_2)$ are numerical
coefficients expected to be of order unity (Ogilvie \& Latter 2013).
The conditions under which this occurs in binary systems were first
studied by Bate et al.~(2000), with the objective of studying circumstellar
discs. They estimated $v'$ from the
internal stress ${\bf G}$ in the disc. More recently, Ogilvie \&
Latter (2013) have presented the first detailed study of the linear
growth of these parametric instabilities, by performing numerical
simulations in the warped shearing sheet formalism. 
For isothermal and quasi-Keplerian discs, they find $A_2\approx 65$ and $A_1\approx 45$
(using $q=1.6$).  These results should be taken with some caution, as
they are only valid for $\psi \lesssim 0.01$ and likely depend on
the vertical profile of the disc.  However, they are probably the best
estimate currently available when trying to determine whether a given disc
profile is susceptible to the growth of parametric instabilities.

From these results, it is clear that in many discs parametric
instabilities can grow even when the small warp condition $\psi \ll 1$
is satisfied.  They could thus play a significant role in the long
term evolution of the disc. However, the consequences of the
growth of these instabilities on the behavior of the disc are not
known at present. One reasonable assumption is that the warp would
then be damped on a timescale comparable to the growth timescale of
the instabilities (Gammie et al.~2000, Bate et al.~2000), in which
case the parametric instabilities would effectively act as a lower
bound on the viscosity parameter $\alpha$ entering the equations for the
propagation of bending waves in the disc. That bound would then depend on the maximum value of $\psi$,
and roughly follow Eq.~(\ref{eq:param}).
If this is the case, the parametric
instabilities could significantly affect the damping timescale of the
warp in many relevant astrophysical configurations (Bate et al.~2000;
see also the following subsections). 

\subsection{Effects on Circumbinary discs}

By combining the local prescription for the growth of the parametric
instability of Ogilvie \& Latter (2013) and our calculation for the
dimensionless warp $\psi$ in circumbinary discs, we can now obtain an
improved estimate of the region of parameter space in which this
instability would grow. If we consider the typical 'flaring disc' model
used in Sec.~\ref{sec:diffprof} and Fig.~\ref{fig:C2} (with $c=0.2$,
$p=1$), we find the maximum dimensionless warp
\be
\psi_{\rm max} \approx 0.09  (4\eta)^2  \left(\frac{2a}{r_{\rm in}}\right)^{4} \left(\frac{0.1}{\delta_{\rm in}}\right)^2 \sin{\beta_{\rm out}}.
\ee
The condition for the growth of parametric instabilities is thus
\be
\sin \beta_{\rm out} \gtrsim 0.07 \left(\frac{\alpha}{0.01}\right)^2 (4\eta)^{-2} \left(\frac{2a}{r_{\rm in}}\right)^{-4} \left(\frac{0.1}{\delta_{\rm in}}\right)^{-2},
\ee
or $\beta_{\rm out} \gtrsim 4^\circ$ for our standard disc parameters.

Ogilvie \& Latter (2013) find that the growth rate of the parametric
instability 
for $\psi$ only $\sim 20\%-50\%$ larger than at the onset of instability is already
$\gtrsim 0.01\Omega$ (for $\alpha\sim 0.005-0.01$). If, as in Bate et al.~(2000) 
we assume that the effect of these instabilities is to damp the inclination of the 
disc on the growth timescale of the instability, we would expect an effective
viscosity of at least $\alpha \sim 0.01$ for any disc inclined by more
than a few degrees.
For standard circumbinary discs, we would then expect 
the inclination to decay to 
$\beta_{\rm out} \ll 1^\circ$ within about $10^5{\rm yrs}$.
After that, the inclination would evolve on a longer (viscous) timescale
(if the effective viscosity is then $\alpha \ll 0.01$).

%

We note that the effects of the local parametric instabilities
on the global effective viscosity $\alpha$ are still speculative, and
that the inclination damping from the growth of the parametric
instabilities remains uncertain. However, as shown in previous section,
even without taking parametric instabilities into account, we
generally get $t_{\rm damp} \ll t_\nu$. Damping of the inclination on
a timescale much shorter than the lifetime of the disc is thus
expected to be the norm even if the parametric instabilities do not
significantly affect the evolution of the inclination.

\subsection{Effects on Circumstellar discs}

We can now perform an identical calculation for circumstellar discs. 
Using a constant thickness profile ($p=1$, $c=0.5$), we get a maximum
dimensionless warp
\be
\psi_{\rm max} \approx 0.02 q^2 \left(\frac{0.1}{\delta}\right)^2  \left(\frac{3r_{\rm out}}{a}\right)^{6} \sin\beta_{\rm out}
\ee
for $\alpha \lesssim 0.025 q (r_{\rm out}/a)^3$ and would predict the growth of parametric instabilities for
\be
\sin\beta_{\rm out} \gtrsim 0.33  q^{-2} \left(\frac{\alpha}{0.01}\right)^2 \left(\frac{0.1}{\delta}\right)^{-2}  \left(\frac{3r_{\rm out}}{a}\right)^{-6}
\ee
or $\beta_{\rm out}\gtrsim 20^\circ$ for our fiducial parameters\footnote{For larger viscosities,
these scalings are modified. But by that point all inclination angles are stable against the growth
of parametric instabilities and inclination damping is dominated by the damping rate given
in Eq.~(\ref{eq:gammacircumstellar})}. The
maximum inclination also generally becomes larger for more asymmetric
binaries or eccentric systems. We can for example consider a system
with $q=1/2$, and choose the outer radius of the disc to be $r_{\rm
  out}=0.4a$ (roughly the expected value for a circular binary,
according to Artymowicz \& Lubow 1994), and find that parametric
instabilities grow for $\beta_{\rm out}\gtrsim 30^\circ$.

In a previous analysis, Bate et al.~(2000)
suggested that parametric instabilities could damp the inclination
of circumstellar discs to within an angle of order $H/r$ 
with respect to the binary orbital plane 
(or $\beta_{\rm out}\sim 5^\circ$ for $\delta = 0.1$) 
on a dynamical time. Their analysis was based on 
a rough estimate of the total internal stress in the disc. 
%
Our calculations based on the local properties of the disc are 
significantly more favorable for the survival of large inclinations
than the result of Bate et al.~(2000).
Although our understanding of parametric instabilities is insufficient to draw
definite conclusions at this point, our results still provide evidence
that even if we take into account parametric instabilities,
circumstellar discs are likely to be able to maintain large
misalignments on timescales comparable to their viscous lifetimes.

\section{KH 15D as a truncated precessing disc}
\label{sec:kh}

In this section, we compare our analytic results to the 1D simulations of
Lodato \& Facchini (2013) (hereafter LF), who presented numerical models
for the KH 15D system. 
The pre-main sequence binary KH 15D shows an unusual light curve,
which has been interpreted as the result of a narrow precessing disc
around the binary (Winn et al.~2004, Chiang \& Murray-Clay 2004).  The
system exhibits deep eclipses occurring at a period of $48.35$days and
lasting $\sim 1$~day.  Radial velocity measurements indicate the
binary component masses of $M_1\sim M_2 \sim 0.5M_\odot$ and orbital
eccentricity $0.68<e<0.80$ (Johnson et al.~2004). The model of Chiang
\& Murray-Clay (2004) gives the disc precession period of 2770~yrs
(with $M_1= M_2 = 0.5M_\odot$ and binary separation $a=0.26{\rm AU}$).

LF studied two numerical models for the KH 15D disc system. Both have the disc 
parameters $r_{\rm in}=4a$ (due to the large eccentricity of the binary), 
$\delta_{\rm in}=0.1$, $\alpha=0.05$, and $c=3/4$. Their surface 
density profiles have index $p=0.5$ and $p=1$, with the corresponding outer disc radii
$x_{\rm out}=6.71$ and $x_{\rm out}=9$, respectively. 
The outer edge of the discs are chosen in order to
recover the desired precession period.

\begin{table}
\caption{ disc models for the KH 15D object. For each parameter, we
  list the values obtained by Lodato \& Facchini (2013) using 1D numerical
  simulations[Num. Sim. (LF)], our numerical results
  obtained by solving for the lowest order eigenmode of the disc [Num. Eigen.], 
  and the results of the approximate analytical formulae obtained in
  Appendix~\ref{App:ShapeFunctions} [Analytical]. The warp for the $p=1/2$ model
  of LF is taken from Fig.2 of LF, not Table I, as the table lists an incorrect 
  value of the warp (S. Facchini, private communication) }
\label{tab:LF}
\begin{tabular}{|c|c|c|c|}
\hline
 & Num. Sim. (LF) & Num. Eigen. & Analytical \\
\hline
$p=1/2$ model & & &\\
\hline
Prec. Period & $2670$yrs & $2800$yrs & $2764$yrs \\ 
\hline
Warp $\delta l_\theta$ & 0.018 & 0.018 & 0.013\\
\hline
$t_{\rm damp}=\gamma^{-1}$ & $4744$yrs & $5704$yrs & $5576$yrs\\
\hline
\hline
$p=1$ model & & &\\
\hline
Prec. Period & $2680$yrs & $2861$yrs & $2769$yrs \\ 
\hline
Warp $\delta l_\theta$ & 0.040 & 0.040 & 0.024\\
\hline
$t_{\rm damp}=\gamma^{-1}$ & $3012$yrs & $3051$yrs & $2918$yrs\\
\hline
\end{tabular}
\end{table}

We first verify that our model for the profile and evolution of the disc
warp agrees with the numerical results obtained by LF. The main
results are listed in Table~\ref{tab:LF}, where we give the precession
period, total warp, and damping timescale obtained by LF,
as well as our results using both the approximate analytical method and the 
explicit numerical computation of the disc eigenmode. We see that the two 
numerical methods agree extremely well, with the worst disagreement being 
on the damping timescale of the $p=0.5$ model ($\sim 20\%$, while other quantities
agree to within a few percents). The approximate analytical model gives
equivalently good results for the precession and damping timescales,
but perform more poorly for 
the total disc warp ($\sim 50\%$ error). The reason for this larger error can 
be easily understood if we also compute the twists, which are
$8^\circ$ and $14^\circ$ for $p=0.5$ and $p=1$ respectively. This means that the
twist modifies $\hatl$ by $15\%-25\%$, while the warp is only a
$2\%-4\%$ effect. Accordingly, the second order terms in the twist, which
we have neglected in the computation of the warp, can significantly affect the warp. 
Note that if we choose 
$\alpha \lesssim 0.01$, then the approximate analytical method agrees with the eigenmode analysis
much better, as shown in Sec.~\ref{sec:numsol}.  

We can now go further and consider the 
potential influence of parametric instabilities. For the disc to
survive multiple precession periods without aligning with the binary,
the viscosity needs to be significantly smaller than $\alpha=0.05$. 
LF estimate that $\alpha < 0.005$ is required,
assuming the expected linear relation $t_{\rm damp} \propto
\alpha^{-1}$. Then we have $\psi_{\rm max}\sim0.009 \sin{\beta}$ for
the $p=0.5$ model and $\psi_{\rm max}\sim0.016 \sin{\beta}$ for the
$p=1$ model. For $\alpha=0.005$, parametric instabilities will thus
develop for $\beta>10^\circ$ (resp. $5^\circ$) for the $p=0.5$
(resp. $p=1$) model.
In the Chiang \& Murray-Clay (2004) models, inclinations of $\sim
10^\circ-20^\circ$ 
are required to explain the observation.
If the parametric instabilities damp the inclination of the disc, 
some fine tuning of the disc parameters is needed to (i) maintain solid body
precession for multiple precession periods, (ii) avoid fast damping due to
parametric instabilities and (iii) maintain a significant disc inclination.
For $\alpha\sim 0.005$, these models remain 
possible, with damping timescales of order of $10-20$ precession
periods.  Considering the uncertainties related to the effects of the
parametric instabilities on the long term evolution of the disc, this
consideration cannot rule out the precessing disc model for KH
15D. But there are certainly mild tensions between that model and the
idea that the warp is damped on the growth timescale of parametric
instabilities.

\section{Discussion and Conclusion}

In this paper, we have presented analytical calculations for the
long-term evolution of inclined/warped accretion discs under the
influence of an axisymmetric perturbing potential. Such evolution is
governed by the combined effects of disc bending waves and viscous
dissipation. Our calculations can be applied to misaligned
circumbinary and circumstellar discs in protostellar systems, as well
as to other astrophysical discs (such as discs formed in neutron star
binary mergers and in tidal disruption events). In particular, we have
derived approximate expressions for the global precession rate and the
disc alignment timescale due to viscous dissipation, as well as 
expressions for the disc warp and internal stress profiles.  Our results build on the
analysis performed by Papaloizou \& Terquem (1995) for inviscid
circumstellar discs, and the formulation of the bending wave equations
with finite viscosity as an eigenvalue problem by Lubow \& Ogilvie
(2000). Our approximate analytic results by-pass the much more
time-consuming simulations of the evolution of disc warps, as well as
numerical calculations of warp eigenmodes. In essence, our analytic
solution corresponds to the lowest-order eigenmode, which dominates
the long-term evolution of the disc warp and inclination. We have
compared our results to previous numerical calculations (when
available) and found good agreement in general.  Our analytic results
are useful for various applications and also provide insights into the
disc warp evolution under various conditions/parameters.

We have applied our results to inclined/warped circumbinary discs.
For the viscosity parameter $\alpha\sim 0.01$ (and typical protoplanetary disc
parameters), we find that the alignment timescale of the disc is comparable to its
global precession timescale, but much shorter than its viscous evolution
timescale. For the inclination to survive for a significant fraction
of the disc lifetime, we need either a high binary eccentricity or an
asymmetric binary, in which case the inner edge of the disc is 
pushed out and the torque acting on the disc is smaller, or an
abnormally thick disc, which is unlikely.  
We have also shown that the density and temperature profiles of the
disc can play a significant role in the magnitude of the disc warping
and the inclination damping timescale, with more strongly `flaring'
discs (i.e. discs with rapidly growing scaleheight $H$) allowing the
inclination to survive for a larger fraction of the disc's lifetime.

Applying our analytical results to circumstellar discs in binaries, we
find that inclined discs tend to have very small warps (and smaller
than circumbinary discs).  Our results are, in this
respect, significantly more favorable for the survival of large inclinations in 
circumstellar discs than previous studies which rely on the global properties 
of the disc to estimate this damping timescale.
Thus, misaligned circumstellar discs are expected to be quite common
in proto-binaries. Qualitatively, the key difference between
circumbinary discs and circumstellar discs is that in former/latter,
the binary torque is exerted at the inner/outer region of the disc,
which contains small/larger amount of angular momentum, leading to
relatively large/small warp and thus faster/slower viscous damping.

We have explored the possibility of efficient damping of disc warp
through parametric instabilities associated with the strong velocity
shears across the disc. Our updated estimate for the alignment
timescale of the disc shows that, for circumbinary discs, the
inclination angle above which parametric instabilities are likely to
drive rapid damping of the inclination is small 
(about $4^\circ$ for typical disc parameters; see Sec.~5.2),
thus making it difficult to maintain inclination angles of 
more than a few degrees
for a significant part of the lifetime of the disc. 
The recent discoveries of circumbinary planetary systems with negligible or
small inclinations (such as Kepler-413; see Kostov et al.~2014) between the
binary plane and the planetary orbital plane are consistent with this 
picture.
On the other hand,
we find that significant misalignments could be possible in
circumstellar discs, even when parametric instabilities are taken into
account. In this respect, our predictions based on the approximate
local prescription for the growth of parametric instabilities derived
by Ogilvie \& Latter (2013) differ from the earlier results of Bate et
al.~(2000), which predicted rapid damping of the inclination to 
less than the disc thickness $H/r$.

We have also showed that the above considerations apply not only to 
nearly aligned discs but also to relatively low mass (nearly) anti-aligned discs, 
except that in the latter case the disc 
may 
counter-align with the binary.  
The smaller inner radius of counter-aligned discs is also likely to make
counter-alignment even faster than alignment. 

As a concrete application of our model, we verified that our calculations
are in agreement with the numerical results of Lodato \&
Facchini (2013), who modeled the precessing circumbinary disc system 
in KH 15D (following an earlier analytical model by Chiang \&
Murray-Clay 2004). Our result for the inclination damping shows that 
there exist some tensions between the required
misalignment angle and the fact that that misalignment should be
maintained for a few precession periods, at least if parametric
instabilities damp the inclination efficiently. In that case, 
a misalignment angle of $\sim 10^\circ$ can only be maintained for
$\sim 10$ precession periods, if the parameters of the system are
tuned appropriately.  This is however sufficient to explain current
observations - and the tensions disappear if the parametric
instabilities 
do not lead to efficient inclination damping.


\section*{Acknowledgments}
This work has been supported in part by the NSF grants AST-1008245,
AST-1211061 and the 
NASA grants NNX12AF85G, NNX14AG94G. 
FF gratefully acknowledges
support from the Vincent and Beatrice Tremaine postdoctoral fellowship,
from the NSERC Canada, from the Canada Research Chairs Program, and
from the Canadian Institute for Advanced Research.


\appendix

\section{Shape functions for power-law disc profiles}
\label{App:ShapeFunctions}

In the case of a power-law disc profile,
\ba
&&\Sigma(r) = \Sigma_{\rm in} \left(\frac{r}{r_{\rm in}}\right)^{-p}, c_s(r) = c_{\rm in} \left(\frac{r}{r_{\rm in}}\right)^{-c},\\
&&Z(r) = Z_{\rm in} \left(\frac{r}{r_{\rm in}}\right)^{-z}, K(r) = K_{\rm in} \left(\frac{r}{r_{\rm in}}\right)^{-k},\\
&&\Omega(r)=\Omega_{\rm in}\left(\frac{r}{r_{\rm in}}\right)^{-3/2}.
\ea
the shape functions $\Gamma$ defined in Sec.~\ref{sec:ana} can be written analytically.
\ba
\Gamma_p &=& \frac{F(1,x_{\rm out};1-p-z)}{F(1,x_{\rm out};5/2-p)}\\
\Gamma_G(x) &=& F(1,x;1-p-z)-\Gamma_p F(1,x;5/2-p)\\
\Gamma_{1}(x) &=&   \Gamma_p \int_x^{x_{\rm out}}dy \,y^{p+2c-3/2} \Gamma_G(y)\\
\Gamma_{2}(x) &=&  \int_x^{x_{\rm out}}dy \,y^{p+2c-k-3} \Gamma_G(y)\\
\Gamma_{3}(x) &=&  \int_x^{x_{\rm out}}dy \,y^{p+2c-3} \Gamma_G(y) \\
\Gamma_I &=&  \frac{\int_1^{x_{\rm out}}dy\,y^{p+2c-3}\Gamma_G^2(y)}{F(1,x_{\rm out};5/2-p)}\\
&=& \frac{\int_1^{x_{\rm out}}dy \left(y^{-p-z}-\Gamma_p y^{3/2-p}\right)\Gamma_3(y)}{F(1,x_{\rm out};5/2-p)}
\ea
where we defined $F(x,y;n)$ as
\ba
F(x,y;a)&=& \frac{y^a-x^a}{a}\, (a\neq 0)\\
&=& \ln{(y/x)}\, (a=0)
\ea
From these equations, we can clearly see that $\Gamma_G(1)=\Gamma_G(x_{\rm out})=0$, and thus $\Gv$ respects the no-torque boundary condition. 
The integral expressions for $\Gamma_{1,2,3}$ are easy to compute, but relatively cumbersome. Under the relatively general
assumption that $p<5/2$ and $p+z>1$, we get
\ba
\Gamma_{1}(x) &=& -\Gamma^2_p
\frac{F(x,x_{\rm out};2+2c)-F(x,x_{\rm out};p+2c-1/2)}{5/2-p}
\nonumber\\
&&+ \Gamma_p
\frac{F(x,x_{\rm out};2c-z+1/2)-F(x,x_{\rm out};p+2c-1/2)}{1-p-z}
\nonumber
\ea

\ba
\Gamma_{2}(x) &=& -\Gamma_p
\frac{F(x,x_{\rm out};1/2+2c-k)-F(x,x_{\rm out};p+2c-2-k)}{5/2-p}
\nonumber\\
&&+
\frac{F(x,x_{\rm out};2c-z-1-k)-F(x,x_{\rm out};p+2c-2-k)}{1-p-z}
\nonumber
\ea

\ba
\Gamma_{3}(x) &=& -\Gamma_p
\frac{F(x,x_{\rm out};1/2+2c)-F(x,x_{\rm out};p+2c-2)}{5/2-p}
\nonumber\\
&&+
\frac{F(x,x_{\rm out};2c-z-1)-F(x,x_{\rm out};p+2c-2)}{1-p-z}.
\nonumber
\ea
Finally, $\Gamma_I$ can be recovered using the identity
\be
\int_1^{x_{\rm out}} dy\,y^{a-1} F(y,x_{\rm out};b) = \frac{F(1,x_{\rm out};a+b)-F(1,x_{\rm out};b)}{a}
\ee
which is valid for $a\neq 0$. We then get
\ba
\Gamma_I &=& \frac{1}{F(1,x_{\rm out};5/2-p)} \bigg(\frac{\Gamma_p \Gamma_3(1;0)}{5/2-p}-\frac{\Gamma_3(1;0)}{1-p-z} \nonumber \\
&&+\frac{\Gamma_3(1;1-p-z)}{1-p-z}-\frac{\Gamma_p \Gamma_3(1;5/2-p)}{5/2-p}\bigg)
\ea
with the notation
\ba
\Gamma_3(x;A) &=& -\Gamma_p
\frac{F(x,x_{\rm out};1/2+2c+A)-F(x,x_{\rm out};p+2c-2+A)}{5/2-p}
\nonumber\\
&&+
\frac{F(x,x_{\rm out};2c-z-1+A)-F(x,x_{\rm out};p+2c-2+A)}{1-p-z}.
\nonumber
\ea

\end{document}